\documentclass[oneside,english,onecolumn]{elsart3p}
\usepackage[T1]{fontenc}
\usepackage[latin1]{inputenc}
\usepackage{amsmath}
\usepackage{graphicx}
\usepackage{amssymb}

\makeatletter

\newcommand{\noun}[1]{\textsc{#1}}

\usepackage{babel}
\makeatother
\begin{document}
\begin{frontmatter}

\title{Efficient calculation of the Coulomb matrix and its expansion around
$\mathbf{k}=\mathbf{0}$ within the FLAPW method}

\author[IFF]{Christoph Friedrich\corauthref{cor}}
\corauth[cor]{Corresponding author.}
\ead{c.friedrich@fz-juelich.de}
\author[Paderborn]{Arno Schindlmayr}
\author[IFF]{Stefan Blügel}
\address[IFF]{Institut für Festkörperforschung and Institute for Advanced Simulation, Forschungszentrum Jülich, 52425 Jülich, Germany}
\address[Paderborn]{Department Physik, Universität Paderborn, 33095 Paderborn, Germany}

\begin{abstract}
We derive formulas for the Coulomb matrix within the full-potential
linearized augmented-plane-wave (FLAPW) method. The Coulomb matrix
is a central ingredient in implementations of many-body perturbation
theory, such as the Hartree-Fock and $GW$ approximations for the
electronic self-energy or the random-phase approximation for the dielectric
function. It is represented in the mixed product basis, which combines
numerical muffin-tin functions and interstitial plane waves constructed
from products of FLAPW basis functions. The interstitial plane waves
are here expanded with the Rayleigh formula. The resulting algorithm
is very efficient in terms of both computational cost and accuracy
and is superior to an implementation with the Fourier transform of
the step function. In order to allow an analytic treatment of the
divergence at $\mathbf{k}=\mathbf{0}$ in reciprocal space, we expand
the Coulomb matrix analytically around this point without resorting
to a projection onto plane waves. Without additional approximations,
we then apply a basis transformation that diagonalizes the Coulomb
matrix and confines the divergence to a single eigenvalue. At the
same time, response matrices like the dielectric function separate
into head, wings, and body with the same mathematical properties as
in a plane-wave basis. As an illustration we apply the formulas to
electron-energy-loss spectra (EELS) for nickel at different $\mathbf{k}$
vectors including \textbf{$\mathbf{k=0}$}. The convergence of the
spectra towards the result at $\mathbf{k=0}$ is clearly seen. Our
all-electron treatment also allows to include transitions from $3s$
and $3p$ core states in the EELS spectrum that give rise to a shallow
peak at high energies and lead to good agreement with experiment.
\end{abstract}
\end{frontmatter}

\section{Introduction}

For the ab initio calculation of electronic excitations and spectroscopic
functions, where variational ground-state schemes like Kohn-Sham density-functional
theory \cite{Hohenberg1964} are not strictly applicable, many-body
perturbation theory has now become the method of choice in applications
to solids and their surfaces. It is based on a Green-function formalism
and an adiabatic switching-on of the Coulomb interaction \cite{Mahan2000}.
In this way the Green function of the fully interacting many-electron
system can be expanded in powers of the Coulomb potential, generating
a series of Feynman diagrams with increasing complexity. Practical
approximations can be designed by terminating the series at a given
order or restricting the summation to certain classes of self-energy
diagrams that describe dominant scattering processes. A prominent
example is the exchange-only Hartree-Fock approximation, which includes
all electronic interaction effects up to linear order in the Coulomb
potential, while additional correlation effects resulting from dynamic
screening in an itinerant electron system are taken into account in
the $GW$ approximation \cite{Hedin1965}. The latter has been successfully
applied to a variety of materials, especially semiconductors, and
generally yields electronic band structures and quasiparticle properties
in very good agreement with experimental data \cite{Aulbur2000}.
The dielectric function, which already appears as an intermediate
quantity in the $GW$ approximation, can be expanded in a similar
manner and is itself the key quantity for the theoretical description
of optical absorption and related spectroscopies.

For a numerical evaluation of self-energies or dielectric functions
in diagrammatic terms it is necessary to project the Coulomb potential,
as well as Green functions and all other relevant propagators, onto
a suitable basis set. Within the diagrammatic expansion the Coulomb
interaction describes the elastic scattering of two electrons or holes,
with a possible momentum transfer between initial and final states.
The basis for the matrix representation of the Coulomb potential must
hence be able to properly describe products of initial-state and final-state
wave functions. So far most practical implementations have employed
a plane-wave basis set in combination with norm-conserving pseudopotentials.
As the product of two plane waves is again a plane wave, this approach
has the advantage that products of wave functions can easily be expressed
in the same basis as the original wave functions themselves. In addition,
fast Fourier transformations may be exploited, and the Coulomb matrix
in reciprocal space is known analytically. For semiconductors, in
particular, sophisticated theoretical calculations of optical absorption
\cite{Albrecht1998} and electron-energy-loss spectra \cite{Olevano2001},
which also include excitonic contributions, have been performed in
this way.

While the plane-wave pseudopotential approach works well for \emph{sp}-bonded
semiconductors and simple metals, it becomes inefficient for transition
metals and rare earths, where a large number of plane waves are needed
to accurately describe the localized \emph{d} or \emph{f} orbitals.
A similar problem occurs in oxides and other compounds involving first-row
elements due to the hard pseudopotentials that only contain minimal
screening of the ionic core by the innermost 1\emph{s} electrons.
Therefore, these materials are best studied within an all-electron
scheme that treats core and valence shells on an equal footing and
already incorporates the rapid oscillations of the wave functions
close to the nuclei in the basis functions themselves. Here we focus
on the full-potential linearized augmented-plane-wave (FLAPW) method
\cite{Singhbook}, which is widely used for electronic-structure calculations
of such materials. It divides space into nonoverlapping muffin-tin
spheres centered at the atomic positions and into the interstitial
region. Inside the muffin-tin spheres the basis functions are constructed
from numerical solutions of the radial Schrödinger equation with fixed
energy parameters, whose products lie outside the vector space spanned
by the original basis functions. Therefore, products of the original
basis functions may instead be used to construct a mixed product basis
\cite{Kotani2002}, in which the matrix elements of the Coulomb potential
with initial and final states are then accurately represented. 

While the Coulomb matrix is diagonal in a plane-wave basis and given
by a simple analytical expression, its evaluation in the mixed product
basis of the FLAPW scheme is much more cumbersome. First, the matrix
is no longer diagonal, and all elements must be calculated numerically.
This requires an efficient computational procedure. Second, due to
the long-range nature of the Coulomb potential $v(r)=1/r$ in real
space, the matrix diverges in the limit of small wave vectors $\mathbf{k}\rightarrow\mathbf{0}$.
Whereas this divergence is confined to the single head element in
the case of a plane-wave basis, all matrix elements now contain divergent
terms proportional to $1/k^{2}$ and $1/k$. Previous all-electron
implementations \cite{Ku2002,Puschnig2002} of many-body perturbation
theory have often bypassed this problem by reverting to a plane-wave
basis for the Coulomb potential and related propagators, such as the
dielectric function, but the projection leads to a loss of accuracy,
because the rapid oscillations of the orbitals close to the atomic
nuclei cannot be resolved then. As a consequence, physical effects
like core polarization are inadequately described. An alternative
approach, the so-called offset-$\Gamma$ method, employs an auxiliary
$\mathbf{k}$-point mesh that is shifted from the origin by a small
but finite amount \cite{Kotani2002,Lebegue2003}. In this way it avoids
the singularity, but the use of additional meshes increases the numerical
cost; even in the most favorable case, for cubic symmetry, the number
of $\mathbf{k}$ points must at least be doubled. Furthermore, the
convergence of Brillouin-zone (BZ) integrals involving the Coulomb
matrix, for example for the $GW$ self-energy, may be slow with respect
to $\mathbf{k}$-point sampling due to the approximate treatment of
the quantitatively important region near the zone center. 

In this work we derive formulas for the Coulomb matrix in the mixed
product basis including its mathematically exact expansion around
$\mathbf{k}=\mathbf{0}$, which involves terms proportional to $1/k^{2}$
and $1/k$ as well as constant terms. A proper treatment of the small-wave-vector
limit is especially important for the theoretical description of optical
spectroscopies with zero momentum transfer, but also for the calculation
of the nonlocal Hartree-Fock potential or the $GW$ self-energy, which
both involve an integration over the BZ. In a second step, to simplify
the numerical treatment we then apply a basis transformation that
diagonalizes the Coulomb matrix. This eliminates all $1/k$ terms
and again restricts the $1/k^{2}$ divergence to a single diagonal
element, belonging to a constant eigenfunction. The final situation
is thus once more analogous to a plane-wave representation, where
the dielectric function naturally decomposes into head, wing, and
body elements, but we retain the full accuracy of the FLAPW basis
set. Furthermore, the present algorithm is very efficient; the computational
time for a well-converged Coulomb matrix with $10^{5}$ elements takes
less than a second on a modern single-CPU personal computer. The present
algorithm is implemented in \noun{Spex} \cite{Spex}, a computer
code for the calculation of excitation spectra and quasiparticle energies
within the $GW$ approximation. 

This paper is organized as follows. Section \ref{sec:Basis-Sets}
shortly describes the FLAPW method and the mixed product basis used
in this work. The formulas for the Coulomb matrix at finite wave vectors
are derived in Section~\ref{sec:Coulomb}. We then discuss its expansion
around $\mathbf{k}=\mathbf{0}$ and the subsequent diagonalization
in Section~\ref{sec:Expansion}. As a practical illustration, in
Section~\ref{sec:test} we present electron-energy-loss spectra of
Ni calculated at finite $\mathbf{k}$ vectors as well as $\mathbf{k}=\mathbf{0}$
within the random-phase approximation. Finally, Section \ref{sec:Summary}
summarizes our main conclusions. Unless stated otherwise we use Hartree
atomic units.

\section{Basis sets\label{sec:Basis-Sets}}

\subsection{FLAPW method}

In the FLAPW method space is divided into nonoverlapping atom-centered
muffin-tin (MT) spheres and the interstitial region (IR). The core-electron
wave functions, which are (mostly) confined to the MT spheres, are
directly obtained from a solution of the fully relativistic Dirac
equation. The valence-electron wave functions with spin $\sigma$
are expanded in interstitial plane waves (IPW) in the interstitial
region and numerical functions $u_{almp}^{\sigma}(\mathbf{r})=u_{alp}^{\sigma}(r)Y_{lm}(\mathbf{e_{r}})$
inside the MT sphere of atom $a$ with position vector $\mathbf{R}_{a}$.
The latter comprise solutions of the Kohn-Sham equation\begin{equation}
\left[-\frac{1}{2}\nabla^{2}+\overline{V}_{\mathrm{eff},a}^{\sigma}(r)\right]u_{alm0}^{\sigma}(\mathbf{r})=\epsilon_{al}^{\sigma}u_{alm0}^{\sigma}(\mathbf{r})\end{equation}
for $p=0$ and their first energy derivatives $u_{alm1}^{\sigma}(\mathbf{r})=\partial u_{alm0}^{\sigma}(\mathbf{r})/\partial\epsilon_{al}^{\sigma}$
for $p=1$, where $\overline{V}_{\mathrm{eff},a}^{\sigma}(r)$ is
the spherical average of the effective potential, $\epsilon_{al}^{\sigma}$
are suitably chosen energy parameters, and $Y_{lm}(\mathbf{e_{r}})$
denote the spherical harmonics. The notation $\mathbf{e}_{\mathbf{r}}=\mathbf{r}/r$
with $r=|\mathbf{r}|$ indicates the unit vector in the direction
of $\mathbf{r}$. In a given unit cell the Kohn-Sham wave function
at a wave vector $\mathbf{k}$ with band index $n$ and spin $\sigma$
is then given by\begin{equation}
\varphi_{n\mathbf{k}}^{\sigma}(\mathbf{r})=\left\{ \begin{array}{ll}
{\displaystyle \frac{1}{\sqrt{N}}\sum_{l=0}^{l_{\mathrm{max}}}\sum_{m=-l}^{l}\sum_{p=0}^{1}}A_{almp}^{n\mathbf{k}\sigma}u_{almp}^{\sigma}(\mathbf{r}-\mathbf{R}_{a}) & \textrm{ if }\mathbf{r}\in\mathrm{MT}(a)\\
{\displaystyle \frac{1}{\sqrt{V}}\sum_{|\mathbf{k+G}|\le G_{\mathrm{max}}}}c_{\mathbf{G}}^{n\mathbf{k}\sigma}e^{i(\mathbf{k}+\mathbf{G})\cdot\mathbf{r}} & \textrm{ if }\mathbf{r}\in\mathrm{IR}\end{array}\right.\label{Eq:FLAPW-basis}\end{equation}
with the crystal volume $V$, the number of unit cells $N$, and cutoff
values $l_{\mathrm{max}}$ and $G_{\mathrm{max}}$. The coefficients
$A_{almp}^{n\mathbf{k}\sigma}$ are determined by the requirement
that the wave function is continuous in value and first radial derivative
at the MT sphere boundaries. If desired, additional local orbitals
\cite{Singh1991} or higher energy derivatives \cite{Friedrich2006}
can also be incorporated by allowing $p\ge2$.

\subsection{Mixed product basis\label{sec:mixedbasis}}

The FLAPW method uses continuous basis functions that are defined
everywhere in space but have a different mathematical representation
in the MT spheres and the IR. For the expansion of wave-function products,
however, it is better to employ two separate sets of functions that
are defined only in one of the spatial regions and zero in the other.
In this way, linear dependences that occur only in one region can
easily be eliminated, which overall leads to a smaller and more efficient
basis. The resulting combined set of functions is called the mixed
product basis.

Inside the MT spheres the mixed product basis must accurately describe
the products \begin{equation}
u_{almp}^{\sigma\,^{{\scriptstyle *}}}(\mathbf{r})u_{al'm'p'}^{\sigma}(\mathbf{r})=u_{alp}^{\sigma}(r)Y_{lm}^{*}(\mathbf{e_{r}})u_{al'p'}^{\sigma}(r)Y_{l'm'}(\mathbf{e_{r}})=\sum_{L=|l-l'|}^{l+l'}\sum_{M=-L}^{L}C_{lml'm'LM}U_{aLP}^{\sigma}(r)Y_{LM}(\mathbf{e_{r}})\,,\end{equation}
which we expand in spherical harmonics with the Gaunt coefficients
\begin{equation}
C_{lml'm'LM}=\int Y_{lm}^{*}(\mathbf{e_{r}})Y_{l'm'}(\mathbf{e_{r}})Y_{LM}^{*}(\mathbf{e_{r}})\, d\Omega\,.\label{eq:Gaunt}\end{equation}
The index $P$ counts the radial product functions $U_{aLP}^{\sigma}(r)=u_{alp}^{\sigma}(r)u_{al'p'}^{\sigma}(r)$
for a given angular quantum number $L$. We emphasize again that,
in general, the latter lie outside the vector space spanned by the
original numerical basis functions $\{ u_{almp}^{\sigma}(\mathbf{r})\}$.
Initially, the set of radial product functions is neither normalized
nor orthogonal and usually has a high degree of (near) linear dependence.
An effective procedure to remove these (near) linear dependences is
to diagonalize the overlap matrix and to retain only those eigenvectors
whose eigenvalues exceed a specified threshold value \cite{Aryasetiawan1994}.
In this way the MT functions become orthonormalized. By using both
spin-up and spin-down products in the construction of the overlap
matrix we make the resulting basis spin-independent. If desired, the
basis set may be reduced further by introducing an additional cutoff
value $L_{\mathrm{max}}$ for the angular quantum number. On the other
hand, it must be supplemented with a constant MT function for each
atom in the unit cell, which is later needed to represent the eigenfunction
that corresponds to the divergent eigenvalue of the Coulomb matrix
in the limit $\mathbf{k}\rightarrow\mathbf{0}$. From the resulting
orthonormal MT functions $M_{aLMP}(\mathbf{r})=M_{aLP}(r)Y_{LM}(\mathbf{e_{r}})$
we formally construct Bloch functions \begin{equation}
M_{aLMP}^{\mathbf{k}}(\mathbf{r})=\frac{1}{\sqrt{N}}\sum_{\mathbf{T}}e^{i\mathbf{k}\cdot(\mathbf{T}+\mathbf{R}_{a})}M_{aLP}(|\mathbf{r}-\mathbf{T}-\mathbf{R}_{a}|)Y_{LM}(\mathbf{e}_{\mathbf{r}-\mathbf{T}-\mathbf{R}_{a}})\,.\label{eq:MT_Bloch}\end{equation}
The sum runs over all lattice translation vectors $\mathbf{T}$, and
$M_{aLP}(r)=0$ if $r$ is larger than the muffin-tin radius $s_{a}$. 

In the IR, since the product of two IPWs equals another IPW, we use
the set \begin{equation}
M_{\mathbf{G}}^{\mathbf{k}}(\mathbf{r})=\frac{1}{\sqrt{V}}e^{i(\mathbf{k+G})\cdot\mathbf{r}}\Theta(\mathbf{r})\label{eq:IPW}\end{equation}
with the step function \begin{equation}
\Theta(\mathbf{r})=\left\{ \begin{array}{ll}
0 & \textrm{ if }\mathbf{r}\in\textrm{MT}\\
1 & \textrm{ if }\mathbf{r}\in\textrm{IR}\end{array}\right.\label{eq:stepfunction}\end{equation}
and a cutoff $G_{\mathrm{max}}^{\prime}\le2G_{\mathrm{max}}$ in reciprocal
space. Together with the MT functions we thus obtain the mixed product
basis $\left\{ M_{I}^{\mathbf{k}}(\mathbf{r})\right\} =\left\{ M_{aLMP}^{\mathbf{k}}(\mathbf{r}),M_{\mathbf{G}}^{\mathbf{k}}(\mathbf{r})\right\} $
for the representation of wave-function products. In contrast to the
MT functions, which were explicitly orthonormalized, the IPWs are
not orthogonal to each other; the elements of their overlap matrix
can be calculated analytically and are given by\begin{equation}
\left\langle M_{\mathbf{G}}^{\mathbf{k}}|M_{\mathbf{G}'}^{\mathbf{k}'}\right\rangle =\delta_{\mathbf{kk}'}O_{\mathbf{GG}'}(\mathbf{k})=\delta_{\mathbf{kk}'}\Theta_{\mathbf{G}-\mathbf{G}'}\,,\end{equation}
where\begin{equation}
\Theta_{\mathbf{G}}=\frac{1}{V}\int e^{-i\mathbf{G}\cdot\mathbf{r}}\Theta(\mathbf{r})\, d^{3}r=\left\{ \begin{array}{ll}
{\displaystyle 1-\frac{4\pi}{3\Omega}\sum_{a}s_{a}^{3}} & \textrm{ for }\mathbf{G}=\mathbf{0}\\
{\displaystyle -\frac{4\pi}{\Omega G^{3}}\sum_{a}e^{-i\mathbf{G\cdot R}_{a}}\left[\sin\left(Gs_{a}\right)-Gs_{a}\,\cos\left(Gs_{a}\right)\right]} & \textrm{ for }\mathbf{G}\neq\mathbf{0}\end{array}\right.\label{eq:Fourier_step}\end{equation}
are the Fourier coefficients of the step function (\ref{eq:stepfunction})
and $\Omega$ denotes the unit-cell volume. We also define a second
set, the biorthogonal set\begin{equation}
\tilde{M}_{I}^{\mathbf{k}}(\mathbf{r})=\sum_{J}O_{JI}^{-1}(\mathbf{k})M_{J}^{\mathbf{k}}(\mathbf{r})\end{equation}
with the overlap matrix $O_{IJ}(\mathbf{k})=\left\langle M_{I}^{\mathbf{k}}|M_{J}^{\mathbf{k}}\right\rangle $.
It fulfills the identities\begin{equation}
\left\langle \tilde{M}_{I}^{\mathbf{k}}\right|\left.M_{J}^{\mathbf{k}}\vphantom{\tilde{M}_{I}^{\mathbf{k}}}\right\rangle =\left\langle M_{I}^{\mathbf{k}}\vphantom{\tilde{M}_{J}^{\mathbf{k}}}\right|\left.\tilde{M}_{J}^{\mathbf{k}}\right\rangle =\delta_{IJ}\quad\textrm{ and \quad}\sum_{I}\left|M_{I}^{\mathbf{k}}\vphantom{\tilde{M}_{I}^{\mathbf{k}}}\right\rangle \left\langle \tilde{M}_{I}^{\mathbf{k}}\right|=\sum_{I}\left|\tilde{M}_{I}^{\mathbf{k}}\right\rangle \left\langle M_{I}^{\mathbf{k}}\vphantom{\tilde{M}_{I}^{\mathbf{k}}}\right|=1\,,\end{equation}
where the completeness relation is only valid in the subspace spanned
by the mixed product basis, however. As the MT functions and the IPWs
are defined in different regions of space and the MT functions are
orthonormal, only the IPWs overlap in a nontrivial way. It should
be noted that the overlap matrix is $\mathbf{k}$-dependent because
the size of the mixed product basis varies for different $\mathbf{k}$
vectors.

For the evaluation of the Coulomb matrix elements we have to find
a numerically tractable expression for the IPWs. A straightforward
approach might employ the Fourier transform of the step function and
rewrite (\ref{eq:IPW}) as a sum over reciprocal lattice vectors\begin{equation}
M_{\mathbf{G}}^{\mathbf{k}}(\mathbf{r})=\lim_{G_{\mathrm{PW}}\rightarrow\infty}\frac{1}{\sqrt{V}}\sum_{|\mathbf{G}'|\le G_{\mathrm{PW}}}\Theta_{\mathbf{G}'}e^{i(\mathbf{k+G+G}')\cdot\mathbf{r}}\,,\label{eq:IPW_Gexpand}\end{equation}
where $G_{\mathrm{PW}}$ is a cutoff radius in reciprocal space, for
which we must of course choose a finite value in practice. Eq.~(\ref{eq:IPW_Gexpand})
has a very simple mathematical structure and is easy to implement.
For example, the calculation of the matrix elements IPW-IPW only involves
Fourier coefficients of the step function $\Theta(\mathbf{r})$ and
the Coulomb interaction $1/r$, which are both known analytically.
As an alternative, we may exploit the Rayleigh expansion\begin{equation}
e^{i\mathbf{k}\cdot\mathbf{r}}=\sum_{l=0}^{\infty}4\pi i^{l}j_{l}(kr)\sum_{m=-l}^{l}Y_{lm}^{*}(\mathbf{e_{k}})Y_{lm}(\mathbf{e_{r}})\label{eq:Rayleigh}\end{equation}
involving the spherical Bessel functions $j_{l}(x)$ in order to subtract
the plane waves inside the MT spheres\begin{equation}
M_{\mathbf{G}}^{\mathbf{k}}(\mathbf{r})=\lim_{l_{\mathrm{PW}}\rightarrow\infty}\frac{1}{\sqrt{V}}\left[e^{i\mathbf{q\cdot r}}-4\pi\sum_{\mathbf{T}}\sum_{a}e^{i\mathbf{q}\cdot(\mathbf{T}+\mathbf{R}_{a})}\theta(s_{a}-r')\sum_{l=0}^{l_{\mathrm{PW}}}i^{l}j_{l}(qr')\sum_{m=-l}^{l}Y_{lm}^{*}(\mathbf{e_{q}})Y_{lm}(\mathbf{e_{r'}})\right]\,,\label{eq:exp_expand}\end{equation}
where we use the abbreviations $\mathbf{q=k+G}$ and $\mathbf{r}'=\mathbf{r}-\mathbf{T}-\mathbf{R}_{a}$,
and $\theta(r)$ denotes the Heaviside function. In a practical implementation
we must use a finite maximal angular momentum $l_{\mathrm{PW}}$,
which thus becomes the relevant convergence parameter. Despite its
more complicated mathematical appearance, we have found that this
representation in fact facilitates a considerably faster numerical
evaluation because of the slow convergence of the step function in
(\ref{eq:IPW_Gexpand}) with respect to the number of Fourier coefficients.
We illustrate this point in Section~\ref{sec:test}. In our subsequent
derivation we hence employ expression (\ref{eq:exp_expand}).

\section{Coulomb matrix at finite $\mathbf{k}$\label{sec:Coulomb}}

In this section we derive the formulas for the computation of the
Coulomb matrix elements\begin{equation}
v_{IJ}(\mathbf{k})=\iint\frac{{M_{I}^{\mathbf{k}}}^{*}(\mathbf{r})M_{J}^{\mathbf{k}}(\mathbf{r}')}{\left|\mathbf{r}-\mathbf{r}'\right|}\, d^{3}r\, d^{3}r'\label{eq:coulomb}\end{equation}
for arbitrary finite wave vectors; the limit $\mathbf{k}\rightarrow\mathbf{0}$
is discussed in Section \ref{sec:Expansion}. Due to the composite
basis set $\left\{ M_{I}^{\mathbf{k}}(\mathbf{r})\right\} $, which
consists of MT functions with $I=(aLMP)$ and IPWs with $I=\mathbf{G}$,
the Coulomb matrix is made of four distinct blocks. As it is Hermitian,
however, the two off-diagonal blocks are complex conjugates of each
other, and thus we have to consider only three blocks explicitly,
which correspond to the combinations MT-MT, MT-IPW, and IPW-IPW. Svane
and Andersen \cite{Svane1986} already examined the matrix elements
MT-MT for finite $\mathbf{k}$ vectors in the context of the linearized
muffin-tin orbital (LMTO) method. In the following we summarize the
derivation in a somewhat different notation and then give the expressions
for the additional matrix elements involving IPWs.

\subsection{MT-MT\label{sub:MT-MT}}

If we insert the Bloch representation (\ref{eq:MT_Bloch}) for the
MT functions in \begin{equation}
v_{aLMP,a'L'M'P'}(\mathbf{k})=\iint\frac{M_{aLMP}^{\mathbf{k}^{\,\,\,{\scriptstyle *}}}(\mathbf{r})M_{a'L'M'P'}^{\mathbf{k}}(\mathbf{r}')}{\left|\mathbf{r}-\mathbf{r}'\right|}\, d^{3}r\, d^{3}r'\,,\label{eq:coulomb_MT-MT}\end{equation}
then the integral can be rewritten as \begin{eqnarray}
\lefteqn{v_{aLMP,a'L'M'P'}(\mathbf{k})=\sum_{\mathbf{T}}e^{i\mathbf{k}\cdot(\mathbf{T}+\mathbf{R}_{aa'})}}\label{eq:coulomb_MT-MT_2}\\
 & \times & \iint\frac{M_{aLP}(r)Y_{LM}^{*}(\mathbf{e_{r}})M_{a'L'P'}(\left|\mathbf{r}'-\mathbf{T}-\mathbf{R}_{aa'}\right|)Y_{L'M'}(\mathbf{e}_{\mathbf{r}'-\mathbf{T}-\mathbf{R}_{aa'}})}{\left|\mathbf{r}-\mathbf{r}'\right|}\, d^{3}r\, d^{3}r'\,,\nonumber \end{eqnarray}
where the difference vector \textbf{$\mathbf{R}_{aa'}=\mathbf{R}_{a'}-\mathbf{R}_{a}$}
points from one MT center to another in the same unit cell. The integrals
in (\ref{eq:coulomb_MT-MT_2}) corresponding to $\mathbf{R}=\mathbf{0}$
and $\mathbf{R}\neq\mathbf{0}$ with $\mathbf{R}=\mathbf{T}+\mathbf{R}_{aa'}$
give rise to two contributions\begin{equation}
v_{aLMP,a'L'M'P'}(\mathbf{k})=\delta_{aa'}v_{aLMP,aL'M'P'}^{(\mathrm{a})}+v_{aLMP,a'L'M'P'}^{(\mathrm{b})}(\mathbf{k})\,,\label{eq:coulomb_MT-MT_3}\end{equation}
which we evaluate separately in the following.

Let us first consider the integral for $\mathbf{R}=\mathbf{0}$. It
can be simplified considerably by inserting the identity \begin{equation}
\frac{1}{\left|\mathbf{r}-\mathbf{r}'\right|}=\sum_{l=0}^{\infty}\frac{4\pi}{2l+1}\frac{r_{<}^{l}}{r_{>}^{l+1}}\sum_{m=-l}^{l}Y_{lm}(\mathbf{e_{r}})Y_{lm}^{*}(\mathbf{e_{r'}})\,,\end{equation}
where $r_{<}$ and $r_{>}$ indicate the smaller and larger value
of $\{ r,r'\}$, respectively. After carrying out the angular integrations
we obtain\begin{eqnarray}
\lefteqn{v_{aLMP,aL'M'P'}^{(\mathrm{a})}}\label{eq:coulomb1_MT-MT}\\
 & = & \delta_{LL'}\delta_{MM'}\frac{4\pi}{2L+1}\int_{0}^{s_{a}}M_{aLP}(r)\left[\frac{1}{r^{L-1}}\int_{0}^{r}r'^{L+2}M_{aLP'}(r')\, dr'+r^{L+2}\int_{r}^{s_{a}}\frac{M_{aLP'}(r')}{r'^{L-1}}\, dr'\right]dr\,.\nonumber \end{eqnarray}
The remaining integrations can be easily performed by standard numerical
techniques on a radial mesh.

For the integrals with $\mathbf{R}\neq\mathbf{0}$ we may formally
define a multipole potential \begin{equation}
\Phi(\mathbf{r})=\int\frac{M_{a'L'P'}(\left|\mathbf{r}'-\mathbf{R}\right|)Y_{L'M'}(\mathbf{e_{r'-R}})}{\left|\mathbf{r}-\mathbf{r}'\right|}d^{3}r'=\frac{4\pi}{2L'+1}\frac{Q_{a'L'P'}}{\left|\mathbf{r}-\mathbf{R}\right|^{L'+1}}Y_{L'M'}(\mathbf{e_{r-R}})\label{eq:potential_MT-MT}\end{equation}
that acts in the first MT sphere as a result of a {}``charge distribution''
$M_{a'L'M'P'}(\mathbf{r}'-\mathbf{R})$ in the second, where $Q_{a'L'P'}$
denotes the multipole moments \begin{equation}
Q_{a'L'P'}=\int_{0}^{s_{a'}}r'^{L'+2}M_{a'L'P'}(r')\, dr'\,.\label{eq:MT_moment}\end{equation}
Using the expansion theorem \cite{Skriver1986,Talman1968}\begin{equation}
\frac{4\pi}{2L'+1}\frac{1}{\left|\mathbf{r}-\mathbf{R}\right|^{L'+1}}Y_{L'M'}(\mathbf{e_{r-R}})=(-1)^{L'+M'}\sum_{l=0}^{\infty}\sum_{m=-l}^{l}c_{L'M',lm}\frac{r^{l}}{R^{L'+l+1}}Y_{lm}(\mathbf{e_{r}})Y_{(L'+l)(m-M')}^{*}(\mathbf{e_{R}})\end{equation}
with the symmetric matrix\begin{eqnarray}
c_{L'M',lm} & = & (-1)^{M'}(4\pi)^{2}\frac{[2(L'+l)-1]!!}{(2L'+1)!!(2l+1)!!}C_{L'M'lm(L'+l)(m-M')}\nonumber \\
 & = & (4\pi)^{3/2}\frac{1}{\sqrt{(2L'+1)(2l+1)\left[2(L'+l)+1\right]}}\sqrt{\frac{(L'+l+m-M')!(L'+l-m+M')!}{(L'+M')!(L'-M')!(l+m)!(l-m)!}}\,,\label{eq:cmat}\end{eqnarray}
 the multipole potential (\ref{eq:potential_MT-MT}) created by a
MT function at $\mathbf{R}$ can then be written in terms of radial
functions and spherical harmonics at the origin. The corresponding
{}``electrostatic interaction energy'' is given by\begin{eqnarray}
\lefteqn{\iint\frac{M_{aLP}(r)Y_{LM}^{*}(\mathbf{e_{r}})M_{a'L'P'}(\left|\mathbf{r}'-\mathbf{R}\right|)Y_{L'M'}(\mathbf{e_{r'-R}})}{\left|\mathbf{r}-\mathbf{r}'\right|}\, d^{3}r\, d^{3}r'}\nonumber \\
 & = & (-1)^{L'+M'}c_{L'M',LM}Q_{aLP}Q_{a'L'P'}\frac{1}{R^{L+L'+1}}Y_{(L+L')(M-M')}^{*}(\mathbf{e_{R}})\,.\end{eqnarray}
After performing the sum over lattice vectors in (\ref{eq:coulomb_MT-MT_2})
we eventually obtain \begin{equation}
v_{aLMP,a'L'M'P'}^{(\mathrm{b})}(\mathbf{k})=(-1)^{L'+M'}e^{i\mathbf{k}\cdot\mathbf{R}_{aa'}}c_{L'M',LM}Q_{a'L'P'}Q_{aLP}S_{(L+L')(M-M')}^{aa'}(\mathbf{k})\label{eq:coulomb2_MT-MT}\end{equation}
 with 

\begin{equation}
S_{lm}^{aa'}(\mathbf{k})=\sum_{\mathbf{T}}e^{i\mathbf{k\cdot T}}\frac{1}{\left|\mathbf{T}+\mathbf{R}_{aa'}\right|^{l+1}}Y_{lm}^{*}(\mathbf{e}_{\mathbf{T}+\mathbf{R}_{aa'}})\,,\label{eq:struct}\end{equation}
where the sum runs over all lattice vectors excluding the case $\mathbf{T}+\mathbf{R}_{aa'}=\mathbf{0}$.
We note that $S_{lm}^{aa'}(\mathbf{k})$ is closely related to the
structure constant defined in the context of the LMTO method \cite{Skriver1986};
however, it is not dimensionless and therefore \emph{not} a constant
of a given crystal structure. For the numerical evaluation of $S_{lm}^{aa'}(\mathbf{k})$
one must apply the Ewald summation technique.

\subsection{MT-IPW\label{sub:MT-IPW}}

For the matrix elements in the off-diagonal block\begin{equation}
v_{aLMP,\mathbf{G}}(\mathbf{k})=\iint\frac{M_{aLMP}^{\mathbf{k}^{\,\,\,{\scriptstyle *}}}(\mathbf{r})M_{\mathbf{G}}^{\mathbf{k}}(\mathbf{r}')}{\left|\mathbf{r}-\mathbf{r}'\right|}\, d^{3}r\, d^{3}r'\end{equation}
we can again introduce a formal {}``charge distribution'' given
by $M_{\mathbf{G}}^{\mathbf{k}}(\mathbf{r}')$ that creates a potential\begin{equation}
\Phi(\mathbf{r})=\int\frac{M_{\mathbf{G}}^{\mathbf{k}}(\mathbf{r}')}{\left|\mathbf{r}-\mathbf{r}'\right|}\, d^{3}r'=\frac{1}{\sqrt{V}}\left(4\pi\frac{e^{i\left(\mathbf{k}+\mathbf{G}\right)\cdot\mathbf{r}}}{\left|\mathbf{k}+\mathbf{G}\right|^{2}}-\int_{\mathrm{MT}}\frac{e^{i\left(\mathbf{k}+\mathbf{G}\right)\cdot\mathbf{r}'}}{\left|\mathbf{r}-\mathbf{r}'\right|}\, d^{3}r'\right)\,,\label{eq:potential_MT-IPW}\end{equation}
where the integral runs over the combined volume of all MT spheres,
cutting out the plane waves inside them. The {}``electrostatic interaction
energy'' arising from the first term in the brackets is given by
\begin{equation}
\frac{4\pi}{\sqrt{V}}\frac{1}{q^{2}}\int M_{aLMP}^{\mathbf{k}^{\,\,\,{\scriptstyle *}}}(\mathbf{r})e^{i\mathbf{q\cdot r}}d^{3}r=\frac{(4\pi)^{2}i^{L}}{\sqrt{\Omega}}Y_{LM}^{*}(\mathbf{e_{q}})e^{i\mathbf{G}\cdot\mathbf{R}_{a}}\frac{1}{q^{2}}\int_{0}^{s_{a}}r^{2}M_{aLP}(r)j_{L}(qr)\, dr\,,\end{equation}
where we have again used the Rayleigh expansion (\ref{eq:Rayleigh})
and the abbreviation $\mathbf{q}=\mathbf{k}+\mathbf{G}$. If the exponential
function in the second term on the right-hand side of (\ref{eq:potential_MT-IPW})
is also replaced by the Rayleigh expansion, then the resulting integrals
are equivalent to those considered in Section \ref{sub:MT-MT} above.
We can hence evaluate them in the same way. The resulting final expression
for the Coulomb matrix element\begin{equation}
v_{aLMP,\mathbf{G}}(\mathbf{k})=v_{aLMP,\mathbf{G}}^{(\mathrm{a})}(\mathbf{k})+v_{aLMP,\mathbf{G}}^{(\mathrm{b})}(\mathbf{k})+v_{aLMP,\mathbf{G}}^{(\mathrm{c})}(\mathbf{k})\end{equation}
consists of three distinct terms, which are given by\begin{subequations}\label{eqall:coulomb_MT-IPW}\begin{eqnarray}
v_{aLMP,\mathbf{G}}^{(\mathrm{a})}(\mathbf{k}) & = & \frac{1}{\sqrt{\Omega}}(4\pi)^{2}i^{L}Y_{LM}^{*}(\mathbf{e_{q}})e^{i\mathbf{G}\cdot\mathbf{R}_{a}}\frac{1}{q^{2}}\int_{0}^{s_{a}}r^{2}M_{aLP}(r)j_{L}(qr)\, dr\,,\label{eq:coulomb1_MT-IPW}\\
v_{aLMP,\mathbf{G}}^{(\mathrm{b})}(\mathbf{k}) & = & -\frac{1}{\sqrt{\Omega}}(4\pi)^{2}i^{L}Y_{LM}^{*}(\mathbf{e_{q}})e^{i\mathbf{G}\cdot\mathbf{R}_{a}}\frac{1}{2L+1}\int_{0}^{s_{a}}M_{aLP}(r)\left[\frac{{\cal I}_{L}(q,r)}{r^{L-1}}+r^{L+2}{\cal J}_{aL}(q,r)\right]dr\,,\label{eq:coulomb2_MT-IPW}\\
v_{aLMP,\mathbf{G}}^{(\mathrm{c})}(\mathbf{k}) & = & -\frac{1}{\sqrt{\Omega}}e^{-i\mathbf{k\cdot R}_{a}}Q_{aLP}\sum_{l'=0}^{l_{\mathrm{PW}}}\sum_{m'=-l}^{l}(-1)^{l'+m'}\sum_{a'}e^{i\mathbf{q\cdot R}_{a'}}c_{l'm',LM}Q_{a'l'm'}^{\mathbf{q}}S_{(L+l')(M-m')}^{aa'}(\mathbf{k})\label{eq:coulomb3_MT-IPW}\end{eqnarray}
\end{subequations} with the multipole moments\begin{equation}
Q_{alm}^{\mathbf{q}}=4\pi i^{l}{\cal I}_{l}(q,s_{a})Y_{lm}^{*}(\mathbf{e_{q}})\label{eq:multipole_IPW}\end{equation}
and the integrals\begin{equation}
{\cal I}_{l}(q,r)=\int_{0}^{r}r'^{l+2}j_{l}(qr')\, dr'\quad\textrm{and}\quad{\cal J}_{al}(q,r)=\int_{r}^{s_{a}}\frac{j_{l}(qr')}{r'^{l-1}}\, dr'\,,\label{eq:integral_sphbes_1}\end{equation}
for which analytic expressions are given in appendix \ref{sec:Integrals-sphbes}.

\subsection{IPW-IPW\label{sub:IPW-IPW}}

The remaining integrals \begin{equation}
v_{\mathbf{GG}'}(\mathbf{k})=\iint\frac{M_{\mathbf{G}}^{\mathbf{k}^{{\scriptstyle *}}}(\mathbf{r})M_{\mathbf{G}'}^{\mathbf{k}}(\mathbf{r}')}{\left|\mathbf{r}-\mathbf{r}'\right|}\, d^{3}r\, d^{3}r'\end{equation}
are evaluated in a similar manner. The subtraction of the plane waves
inside the MT spheres now leads to a decomposition of the matrix elements
into four terms\begin{equation}
v_{\mathbf{GG}'}(\mathbf{k})=v_{\mathbf{GG}'}^{(\mathrm{a})}(\mathbf{k})-v_{\mathbf{GG}'}^{(\mathrm{b})}(\mathbf{k})-v_{\mathbf{GG}'}^{(\mathrm{c})}(\mathbf{k})+v_{\mathbf{GG}'}^{(\mathrm{d})}(\mathbf{k})\,.\label{eq:coulomb_IPW-IPW}\end{equation}
The first three can be calculated analytically and yield\begin{subequations}\label{eqall:coulomb1_IPW-IPW}\begin{eqnarray}
v_{\mathbf{GG}'}^{(\mathrm{a})}(\mathbf{k}) & = & \frac{1}{V}\int d^{3}r\,\int d^{3}r'\,\frac{e^{-i(\mathbf{k+G})\cdot\mathbf{r}}e^{i(\mathbf{k+G}')\cdot\mathbf{r}'}}{\left|\mathbf{r}-\mathbf{r}'\right|}=\delta_{\mathbf{GG}'}\frac{4\pi}{\left|\mathbf{k}+\mathbf{G}\right|^{2}}\quad,\\
v_{\mathbf{GG}'}^{(\mathrm{b})}(\mathbf{k}) & = & \frac{1}{V}\int_{\mathrm{MT}}d^{3}r\,\int d^{3}r'\,\frac{e^{-i(\mathbf{k+G})\cdot\mathbf{r}}e^{i(\mathbf{k+G}')\cdot\mathbf{r}'}}{\left|\mathbf{r}-\mathbf{r}'\right|}=(\delta_{\mathbf{GG}'}-\Theta_{\mathbf{G}-\mathbf{G}'})\frac{4\pi}{\left|\mathbf{k}+\mathbf{G}'\right|^{2}}\quad,\\
v_{\mathbf{GG}'}^{(\mathrm{c})}(\mathbf{k}) & = & \frac{1}{V}\int d^{3}r\,\int_{\mathrm{MT}}d^{3}r'\,\frac{e^{-i(\mathbf{k+G})\cdot\mathbf{r}}e^{i(\mathbf{k+G}')\cdot\mathbf{r}'}}{\left|\mathbf{r}-\mathbf{r}'\right|}=(\delta_{\mathbf{GG}'}-\Theta_{\mathbf{G}-\mathbf{G}'})\frac{4\pi}{\left|\mathbf{k}+\mathbf{G}\right|^{2}}\quad,\end{eqnarray}
while we evaluate the fourth term\begin{equation}
v_{\mathbf{GG}'}^{(\mathrm{d})}(\mathbf{k})=\frac{1}{V}\int_{\mathrm{MT}}d^{3}r\,\int_{\mathrm{MT}}d^{3}r'\,\frac{e^{-i(\mathbf{k+G})\cdot\mathbf{r}}e^{i(\mathbf{k+G}')\cdot\mathbf{r}'}}{\left|\mathbf{r}-\mathbf{r}'\right|}\label{eq:coulomb_IPW-IPW_term4}\end{equation}
\end{subequations}by again replacing the exponential functions with
the Rayleigh expansion (\ref{eq:Rayleigh}) and following the procedure
outlined in Section \ref{sub:MT-MT} above. The subsequent summation
over MT spheres and angular momenta yields\begin{eqnarray}
v_{\mathbf{GG}'}^{(\mathrm{d})}(\mathbf{k}) & = & \frac{1}{\Omega}\biggl(\sum_{a}e^{i(\mathbf{G}'-\mathbf{G})\cdot\mathbf{R}_{a}}\sum_{l=0}^{l_{\mathrm{PW}}}\sum_{m}\frac{(4\pi)^{3}}{2l+1}Y_{lm}(\mathbf{e_{q}})Y_{lm}^{*}(\mathbf{e_{q'}}){\cal K}_{al}(q,q')\label{eq:coulomb2_IPW-IPW}\\
 &  & \,\,\,\,\,\,\,\,\,+\sum_{l=0}^{l_{\mathrm{PW}}}\sum_{m=-l}^{l}\,\,\sum_{l'=0}^{l_{\mathrm{PW}}}\sum_{m'=-l}^{l}(-1)^{l'+m'}\sum_{a,a'}e^{-i\mathbf{q\cdot R}_{a}}e^{i\mathbf{q}'\cdot\mathbf{R}_{a'}}c_{l'm',lm}Q_{alm}^{\mathbf{q}*}Q_{a'l'm'}^{\mathbf{q}'}S_{(l+l')(m-m')}^{aa'}(\mathbf{k})\biggr)\nonumber \end{eqnarray}
 with $\mathbf{q}=\mathbf{k}+\mathbf{G}$, $\mathbf{q}'=\mathbf{k}+\mathbf{G}'$
and the double integral \begin{equation}
{\cal K}_{al}(q,q')=\int_{0}^{s_{a}}\int_{0}^{s_{a}}r^{2}r'^{2}j_{l}(qr)j_{l}(q'r')\frac{r_{<}^{l}}{r_{>}^{l+1}}\, dr\, dr'\,.\label{eq:integral_sphbes_2}\end{equation}
For the latter an analytic formula is derived in appendix \ref{sec:Integrals-sphbes}.

\section{Expansion around $\mathbf{k}=\mathbf{0}$\label{sec:Expansion}}

Due to the long-range nature of the Coulomb interaction $v(r)=1/r$
in real space, its Fourier transform $4\pi/k^{2}$ diverges for $\mathbf{k}\rightarrow\mathbf{0}$.
As a consequence, the Coulomb matrix in the mixed product basis also
diverges with a leading term proportional to \textbf{$1/k^{2}$.}
Since the MT functions contain nontrivial $\mathbf{k}$-dependent
coefficients, we further have additional terms proportional to $1/k$.
It is helpful to identify all relevant terms in advance. For this
purpose we formally represent the basis functions by their Fourier
transforms \begin{equation}
M_{I}^{\mathbf{k}}(\mathbf{r})=\frac{1}{\sqrt{V}}\sum_{\mathbf{G}}c_{I\mathbf{G}}(\mathbf{k})e^{i(\mathbf{k+G})\cdot\mathbf{r}}\end{equation}
with the coefficients\begin{equation}
c_{I\mathbf{G}}(\mathbf{k})=\frac{1}{\sqrt{V}}\int e^{-i(\mathbf{k+G})\cdot\mathbf{r}}M_{I}^{\mathbf{k}}(\mathbf{r})\, d^{3}r\,.\label{eq:Fourier_MB}\end{equation}
The sum runs over all reciprocal lattice vectors $\mathbf{G}$. For
the IPWs the coefficients are $\mathbf{k}$-independent and equal
$c_{\mathbf{G}'\mathbf{G}}(\mathbf{k})=\Theta_{\mathbf{G}-\mathbf{G}'}$
for $M_{\mathbf{G}'}^{\mathbf{k}}(\mathbf{r})$, but for the MT functions
they exhibit a nontrivial \textbf{$\mathbf{k}$} dependence. Using
the expansion\begin{equation}
M_{I}^{\mathbf{k}}(\mathbf{r})\sim\frac{1}{\sqrt{V}}\sum_{\mathbf{G}}\left(c_{I\mathbf{G}}+\mathbf{k}\cdot\nabla c_{I\mathbf{G}}+\frac{1}{2}\mathbf{k}^{\mathrm{T}}\Delta c_{I\mathbf{G}}\mathbf{k}\right)e^{i(\mathbf{k+G})\cdot\mathbf{r}}\end{equation}
for $\mathbf{k}\rightarrow\mathbf{0}$ with $c_{I\mathbf{G}}=\left.c_{I\mathbf{G}}(\mathbf{k})\right|_{\mathbf{k}=\mathbf{0}}$,
$\nabla c_{I\mathbf{G}}=\left.\nabla_{\mathbf{k}}c_{I\mathbf{G}}(\mathbf{k})\right|_{\mathbf{k}=\mathbf{0}}$
and $\Delta c_{I\mathbf{G}}=\left.\nabla_{\mathbf{k}}\nabla_{\mathbf{k}}^{\mathrm{T}}c_{I\mathbf{G}}(\mathbf{k})\right|_{\mathbf{k}=\mathbf{0}}$,
we can write the Coulomb matrix elements in this limit as\begin{eqnarray}
v_{IJ}(\mathbf{k}) & \sim & c_{I\mathbf{0}}^{*}c_{J\mathbf{0}}\frac{4\pi}{k^{2}}+\left[c_{I\mathbf{0}}^{*}(\mathbf{e_{k}}\cdot\nabla c_{J\mathbf{0}})+(\mathbf{e_{k}}\cdot\nabla c_{I\mathbf{0}}^{*})c_{J\mathbf{0}}\right]\frac{4\pi}{k}+\left[\left(\mathbf{e_{k}}\cdot\nabla c_{I\mathbf{0}}^{*}\right)\left(\mathbf{e_{k}}\cdot\nabla c_{J\mathbf{0}}\right)\right.\nonumber \\
 &  & \left.+\frac{1}{2}c_{I\mathbf{0}}^{*}(\mathbf{e}_{\mathbf{k}}^{\mathrm{T}}\Delta c_{J\mathbf{0}}\mathbf{e_{k}})+\frac{1}{2}(\mathbf{e}_{\mathbf{k}}^{\mathrm{T}}\Delta c_{I\mathbf{0}}^{*}\mathbf{e_{k}})c_{J\mathbf{0}}\right]4\pi+\sum_{\mathbf{G}\ne\mathbf{0}}c_{I\mathbf{G}}^{*}c_{J\mathbf{G}}\frac{4\pi}{\left|\mathbf{G}\right|^{2}}\,.\label{eq:coulomb_expand_PW}\end{eqnarray}
Evidently, all matrix elements contain divergent contributions proportional
to $1/k^{2}$ in addition to a constant term. Furthermore, if $c_{I\mathbf{G}}(\mathbf{k})$
or $c_{J\mathbf{G}}(\mathbf{k})$ are truly $\mathbf{k}$-dependent,
i.e., for matrix elements that involve MT functions, we also have
terms proportional to $Y_{1m}^{*}(\mathbf{e_{k}})/k$ and $Y_{2m}^{*}(\mathbf{e_{k}})$
arising from the first and second square bracket, respectively (compare
(\ref{eq:kG_Y1Y1}) and (\ref{eq:kG2_})). As a consequence, we can
write\begin{equation}
v_{IJ}(\mathbf{k})\sim v_{IJ}^{(0)}+\sum_{l=0}^{2}\sum_{m=-l}^{l}v_{IJ,lm}^{(1)}\frac{Y_{lm}^{*}(\mathbf{e_{k}})}{k^{2-l}}\,,\label{eq:coulomb_expand}\end{equation}
and from (\ref{eq:coulomb}) follows\begin{equation}
v_{JI}^{(0)}=v_{IJ}^{(0)*}\quad\textrm{and}\quad v_{JI,lm}^{(1)}=(-1)^{m}v_{IJ,l(-m)}^{(1)*}\,.\end{equation}
We will see in Section \ref{sub:Diagonalization} below that the terms
corresponding to $l>0$ can in fact be eliminated if we perform a
basis transformation to the set of eigenvectors of the Coulomb matrix.
Nevertheless, for the sake of completeness we will here give the appropriate
formulas for $v_{IJ,lm}^{(1)}$ with $l>0$ in the original mixed
product basis as well. As in the previous section, we proceed by discussing
the blocks MT-MT, MT-IPW and IPW-IPW individually.

\subsection{MT-MT}

The second term on the right-hand side of (\ref{eq:coulomb_MT-MT_3}),
explicitly given in (\ref{eq:coulomb2_MT-MT}), diverges for $\mathbf{k}\rightarrow\mathbf{0}$
and $L+L'<2$, because the leading term of $S_{lm}^{aa'}(\mathbf{k})$
is proportional to \textbf{$k^{2-l}$}, which is seen in the following
way: For small $\mathbf{k}$ the sum over $\mathbf{T}$ \textbf{}in
(\ref{eq:struct}) is dominated by contributions belonging to large
lattice vectors. Then one can approximate the sum by an integral\begin{equation}
S_{lm}^{aa'}(\mathbf{k})\sim e^{-i\mathbf{k}\mathbf{\cdot R}_{aa'}}\frac{1}{\Omega}\int\frac{e^{i\mathbf{k\cdot T}}}{T^{l+1}}Y_{lm}^{*}(\mathbf{e}_{\mathbf{T}})\, d^{3}T=\frac{4\pi i^{l}}{(2l-1)!!\Omega}e^{-i\mathbf{k}\cdot\mathbf{R}_{aa'}}Y_{lm}^{*}(\mathbf{e}_{\mathbf{k}})k^{l-2}\,,\label{eq:struct1_expand}\end{equation}
where we have used (\ref{eq:Rayleigh}), (\ref{eq:sphbes_prim2}),
and (\ref{eq:sphes_expand}). The same expression appears in the first
term of the reciprocal-space sum corresponding to $\mathbf{G}=\mathbf{0}$
in the Ewald summation. The remaining terms and the real-space sum
yield an additional constant term $S_{lm}^{aa'}$, so that we obtain
\begin{equation}
S_{lm}^{aa'}(\mathbf{k})\sim\frac{4\pi i^{l}}{(2l-1)!!\Omega}e^{-i\mathbf{k}\cdot\mathbf{R}_{aa'}}Y_{lm}^{*}(\mathbf{e_{k}})k^{l-2}+S_{lm}^{aa'}\label{eq:struct_expand}\end{equation}
for $l\le2$. After inserting this expansion into (\ref{eq:coulomb2_MT-MT})
we obtain\begin{eqnarray}
v_{aLMP,a'L'M'P'}^{(0)} & = & \delta_{aa'}v_{aLMP,a'L'M'P'}^{(\mathrm{a})}+(-1)^{L'+M'}c_{L'M',LM}Q_{a'L'P'}Q_{aLP}S_{(L+L')(M-M')}^{aa'}\,,\\
v_{aLMP,a'L'M'P';lm}^{(1)} & = & \delta_{l,L+L'}\delta_{m,M-M'}(-1)^{L'+M'}\frac{4\pi i^{l}}{(2l-1)!!\Omega}c_{L'M',LM}Q_{a'L'P'}Q_{aLP}\end{eqnarray}
with $l\le2$.

\subsection{MT-IPW}

The case MT-IPW is more complicated, because higher orders in the
multipole moments $Q_{a'l'm'}^{\mathbf{k}+\mathbf{G}}$ must be taken
into account when multiplying with the divergent $S_{(L+l')(M-m')}^{aa'}(\mathbf{k})$
in (\ref{eq:coulomb3_MT-IPW}). In particular, we need the expansions
of $Q_{a'00}^{\mathbf{k}+\mathbf{G}}$ ($Q_{a'1m'}^{\mathbf{k}+\mathbf{G}}$)
up to second (first) order\begin{subequations}\label{eqall:moment_expand}\begin{eqnarray}
Q_{a00}^{\mathbf{k}+\mathbf{G}} & \sim & \sqrt{4\pi}\frac{s_{a}^{2}}{G}\left[j_{1}(Gs_{a})-j_{2}(Gs_{a})s_{a}k\,\mathbf{e_{k}\cdot e_{G}}+\frac{1}{2}j_{3}(Gs_{a})s_{a}^{2}k^{2}\left(\mathbf{e_{k}\cdot e_{G}}\right)^{2}-\frac{1}{2}\frac{j_{2}(Gs_{a})}{G}s_{a}k^{2}\right]\,,\nonumber \\
\\Q_{a1m}^{\mathbf{k}+\mathbf{G}} & \sim & 4\pi i\frac{s_{a}^{3}}{G}\left[Y_{1m}^{*}(\mathbf{e_{G}})\left(j_{2}(Gs_{a})-j_{3}(Gs_{a})s_{a}k\,\mathbf{e_{k}\cdot e_{G}}\right)+Y_{1m}^{*}(\mathbf{e_{k}})\frac{j_{2}(Gs_{a})}{G}k\right]\,.\end{eqnarray}
\end{subequations}For $\mathbf{G}=\mathbf{0}$ these simplify to\begin{subequations}\label{eqall:moment_expand2}\begin{eqnarray}
Q_{a00}^{\mathbf{k}} & \sim & \frac{\sqrt{4\pi}}{3}s_{a}^{3}\left(1-\frac{1}{10}s_{a}^{2}k^{2}\right)\,,\\
Q_{a1m}^{\mathbf{k}} & \sim & \frac{4\pi i}{15}Y_{1m}^{*}(\mathbf{e_{k}})s_{a}^{5}k\,.\end{eqnarray}
\end{subequations} Here we have used the identities (\ref{eq:recursion_bessel})--(\ref{eq:Y1_expand}).
In addition, (\ref{eq:coulomb1_MT-IPW}) contributes to $v_{aLMP,\mathbf{G}}^{(1)}$
if $\mathbf{G=0}$. The final expression for $v_{aLMP,\mathbf{G}}^{(0)}$
and $v_{aLMP,\mathbf{G}}^{(1)}$ is written as\begin{eqnarray}
v_{aLMP,\mathbf{G}}^{(0)} & = & v_{aLMP,\mathbf{G}}^{(0\mathrm{a})}+v_{aLMP,\mathbf{G}}^{(0\mathrm{b})}\,,\end{eqnarray}
where the quantity\begin{eqnarray}
v_{aLMP,\mathbf{G}}^{(0\mathrm{a})} & = & (1-\delta_{\mathbf{G0}})v_{aLMP,\mathbf{G}}^{(\mathrm{a})}(\mathbf{0})+v_{aLMP,\mathbf{G}}^{(\mathrm{b})}(\mathbf{0})\\
 &  & -\frac{1}{\sqrt{\Omega}}Q_{aLP}\sum_{l'=0}^{l_{\mathrm{PW}}}\sum_{m'=-l'}^{l'}(-1)^{l'+m'}\sum_{a'}e^{i\mathbf{G\cdot R}_{a'}}c_{l'm',LM}Q_{a'l'm'}^{\mathbf{G}}S_{(L+l')(M-m')}^{aa'}\nonumber \end{eqnarray}
is directly obtained after replacing $S_{lm}^{aa'}(\mathbf{k})$ by
the terms of zeroth order in the expansion (\ref{eq:struct_expand}).
The second term $v_{aLMP,\mathbf{G}}^{(0\mathrm{b})}$ results from
multiplying the divergent terms in (\ref{eq:struct_expand}) with
the higher orders of (\ref{eqall:moment_expand}) in (\ref{eq:coulomb3_MT-IPW})
as well as, in the case $\mathbf{G}=\mathbf{0}$, the term $1/q^{2}$
with the higher orders of $j_{l}(qr)$ in (\ref{eq:coulomb1_MT-IPW}).
After some algebra we obtain\begin{equation}
v_{aLMP,\mathbf{G}}^{(0\mathrm{b})}=\left\{ \begin{array}{ll}
-\frac{(4\pi)^{5/2}}{\Omega^{3/2}}Q_{a0P}\sum_{a'}e^{i\mathbf{G\cdot R}_{a'}}\frac{s_{a'}^{3}}{G}\left[\frac{j_{2}(Gs_{a'})}{2G}-\frac{j_{3}(Gs_{a'})s_{a'}}{6}\right] & \textrm{if }\mathbf{G}\ne\mathbf{0}\textrm{ and }L=0\,,\\
-\frac{(4\pi)^{5/2}}{30\Omega^{3/2}}Q_{a0P}\sum_{a'}s_{a'}^{5}+\frac{(4\pi)^{3/2}}{6\sqrt{\Omega}}\int_{0}^{s_{a}}r^{4}M_{a0P}(r)dr & \textrm{if }\mathbf{G}=\mathbf{0}\textrm{ and }L=0\,,\\
-\frac{(4\pi)^{5/2}}{\Omega^{3/2}}Q_{a0P}\sum_{a'}e^{i\mathbf{G\cdot R}_{a'}}\frac{s_{a'}^{3}}{G}\left[\frac{j_{2}(Gs_{a'})}{2G}-\frac{j_{3}(Gs_{a'})s_{a'}}{6}\right] & \textrm{if }\mathbf{G}\neq\mathbf{0}\textrm{ and }L=1\,,\\
0 & \textrm{otherwise ,}\end{array}\right.\label{eqall:coulomb_expand_corr}\end{equation}
as well as\begin{equation}
v_{aLMP,\mathbf{G};lm}^{(1)}=\delta_{Ll}\delta_{Mm}\left[-\frac{(4\pi)^{5/2}i^{L}}{(2L+1)!!\Omega^{3/2}}Q_{aLP}\sum_{a'}e^{i\mathbf{G\cdot R}_{a'}}Q_{a'00}^{\mathbf{G}}+\delta_{\mathbf{G0}}\frac{(4\pi)^{2}i^{L}Q_{aLP}}{(2L+1)!!\Omega^{1/2}}\right]\,.\end{equation}

\subsection{IPW-IPW}

Finally, in the calculation of the IPW-IPW matrix elements we can
use the fact that the square brackets in (\ref{eq:coulomb_expand_PW})
vanish. This simplifies the derivation considerably, because all angular-dependent
contributions can be discarded from the outset, and hence we have
$v_{\mathbf{GG}',lm}^{(1)}=0$ for $l>0$. We again write\[
v_{\mathbf{GG}'}^{(0)}=v_{\mathbf{GG'}}^{(0\mathrm{a})}+v_{\mathbf{GG}'}^{(0\mathrm{b})}\,,\]
where the first contribution $v_{\mathbf{GG}'}^{(0\mathrm{a})}$ is
given by the nondivergent terms in (\ref{eq:coulomb_IPW-IPW}) after
replacing $S_{lm}^{aa'}(\mathbf{k})$ by $S_{lm}^{aa'}$, which yields\begin{eqnarray}
v_{\mathbf{GG}'}^{(0\mathrm{a})} & = & (1-\delta_{\mathbf{G0}})\left[v_{\mathbf{GG}'}^{(\mathrm{a})}(\mathbf{0})-v_{\mathbf{GG}'}^{(\mathrm{c})}(\mathbf{0})\right]-(1-\delta_{\mathbf{G'0}})v_{\mathbf{GG'}}^{(\mathrm{b})}(\mathbf{0})\nonumber \\
 & + & \frac{1}{\Omega}\biggl(\sum_{a}e^{i(\mathbf{G}'-\mathbf{G})\cdot\mathbf{R}_{a}}\sum_{l=0}^{l_{\mathrm{PW}}}\sum_{m}\frac{(4\pi)^{3}}{2l+1}Y_{lm}(\mathbf{e_{G}})Y_{lm}^{*}(\mathbf{e_{G'}}){\cal K}_{al}(q,q')\\
 &  & \,\,\,\,\,\,\,\,\,+\sum_{l=0}^{l_{\mathrm{PW}}}\sum_{m=-l}^{l}\,\,\sum_{l'=0}^{l_{\mathrm{PW}}}\sum_{m'=-l}^{l}(-1)^{l'+m'}\sum_{a,a'}e^{-i\mathbf{G\cdot R}_{a}}e^{i\mathbf{G}'\cdot\mathbf{R}_{a'}}c_{l'm',lm}Q_{alm}^{\mathbf{G}^{{\scriptstyle *}}}Q_{a'l'm'}^{\mathbf{G}'}S_{(l+l')(m-m')}^{aa'}\biggr)\,.\nonumber \end{eqnarray}
 Further, by inserting the expansions (\ref{eqall:moment_expand})
and (\ref{eqall:moment_expand2}) as well as the $\mathbf{k}$-dependent
terms of (\ref{eq:struct_expand}) into (\ref{eq:coulomb2_IPW-IPW})
one obtains another constant contribution \begin{equation}
v_{\mathbf{GG}'}^{(0\mathrm{b})}=\left\{ \begin{array}{ll}
\lefteqn{\frac{(4\pi)^{3}}{\Omega^{2}}\sum_{a,a'}e^{-i\mathbf{G\cdot R}_{a}}e^{i\mathbf{G}'\cdot\mathbf{R}_{a'}}\frac{s_{a}^{2}s_{a'}^{2}}{GG'}\left\{ -\frac{1}{3}j_{2}(Gs_{a})j_{2}(G's_{a'})s_{a}s_{a'}(\mathbf{e_{G}\cdot e_{G'}})-\frac{1}{6}j_{1}(Gs_{a})j_{3}(G's_{a'})s_{a'}^{2}\right.}\\
\left.\quad-\frac{1}{6}j_{3}(Gs_{a})j_{1}(G's_{a'})s_{a}^{2}+\frac{j_{1}(Gs_{a})j_{2}(G's_{a'})s_{a'}}{2G'}+\frac{j_{2}(Gs_{a})j_{1}(G's_{a'})s_{a}}{2G}\right\}  & \textrm{if }\mathbf{G}\ne\mathbf{0}\textrm{ and }\mathbf{G}'\ne\mathbf{0},\\
\frac{(4\pi)^{3}}{\Omega^{2}}\sum_{a,a'}e^{i\mathbf{G}'\cdot\mathbf{R}_{a'}}\frac{s_{a}^{3}s_{a'}^{2}}{G'}\left\{ \frac{s_{a}^{2}}{30}j_{1}(G's_{a'})-\frac{1}{18}j_{3}(G's_{a'})s_{a'}^{2}+\frac{1}{6}\frac{j_{2}(G's_{a'})}{G'}s_{a'}\right\}  & \textrm{if }\mathbf{G}=\mathbf{0}\textrm{ and }\mathbf{G}'\ne\mathbf{0},\\
\frac{(4\pi)^{3}}{\Omega^{2}}\sum_{a,a'}e^{-i\mathbf{G}\cdot\mathbf{R}_{a'}}\frac{s_{a}^{3}s_{a'}^{2}}{G}\left\{ \frac{s_{a}^{2}}{30}j_{1}(Gs_{a'})-\frac{1}{18}j_{3}(Gs_{a'})s_{a'}^{2}+\frac{1}{6}\frac{j_{2}(Gs_{a'})}{G}s_{a'}\right\}  & \textrm{if }\mathbf{G}\neq\mathbf{0}\textrm{ and }\mathbf{G}'=\mathbf{0},\\
\frac{(4\pi)^{3}}{90\Omega^{2}}\sum_{a,a'}s_{a}^{3}s_{a'}^{3}\left(s_{a}^{2}+s_{a'}^{2}\right) & \textrm{if }\mathbf{G}=\mathbf{G}'=\mathbf{0}\,.\end{array}\right.\end{equation}
For the calculation of $v_{\mathbf{GG}',00}^{(1)}$ we must take the
divergent terms of (\ref{eqall:coulomb1_IPW-IPW}) into account and
eventually obtain\begin{equation}
v_{\mathbf{GG}',00}^{(1)}=\frac{(4\pi)^{5/2}}{\Omega^{2}}\sum_{a,a'}e^{-i\mathbf{G\cdot R}_{a}}e^{i\mathbf{G}'\cdot\mathbf{R}_{a'}}Q_{a00}^{\mathbf{G}^{\,{\scriptstyle *}}}Q_{a'00}^{\mathbf{G}'}+(4\pi)^{3/2}\left[\Theta_{\mathbf{G-G}'}(\delta_{\mathbf{G0}}+\delta_{\mathbf{G}'\mathbf{0}})-\delta_{\mathbf{G0}}\delta_{\mathbf{G}'\mathbf{0}}\right]\,.\end{equation}

\subsection{Diagonalization\label{sub:Diagonalization}}

In a pure plane-wave representation response matrices and similar
quantities decompose into head $\chi_{\mathbf{00}}$, wings $\chi_{\mathbf{G0}}$,
$\chi_{\mathbf{0G}'}$, and body $\chi_{\mathbf{GG}'}$ with $\mathbf{G},\mathbf{G}'\neq\mathbf{0}$.
These behave differently for $\mathbf{k}\rightarrow\mathbf{0}$. For
the density response function, as an example, head and wing elements
are quadratic and linear in \textbf{$k$}, respectively, while the
body elements remain finite but still exhibit an angular $\mathbf{k}$
dependence. As the mixed product basis is related to the set of plane
waves by means of a basis transformation, these matrix elements will
now mix in a complicated manner. It is hence desirable to make another
transformation that restores the convenient mathematical properties
of the plane-wave basis. For this purpose the basis must include a
constant function, which corresponds to the limit of $e^{i\mathbf{k\cdot r}}/\sqrt{V}$
for $\mathbf{k\rightarrow0}$ and is responsible for the decomposition
into head, wings, and body. Such a basis is given by the functions\begin{equation}
E_{\mu}^{\mathbf{k}}(\mathbf{r})=\sum_{I}E_{\mu I}^{\mathbf{k}}M_{I}^{\mathbf{k}}(\mathbf{r})\,,\label{eq:basis_eigenvec}\end{equation}
where $E_{\mu I}^{\mathbf{k}}$ is the $I$th component of the $\mu$th
eigenvector of $v_{IJ}(\mathbf{k})$. In this basis the Coulomb matrix
$v_{\mu\nu}(\mathbf{k})$ becomes diagonal, which is also advantageous
in matrix multiplications involving $v_{\mu\nu}(\mathbf{k})$, e.g.,
for the calculation of the dielectric function. Furthermore, the eigenvalues
are a direct measure for the probability of the elastic scattering
between two particles. Small eigenvalues thus identify less important
scattering processes that might be neglected, leading to a smaller
and optimized basis set after removal of the corresponding eigenvectors. 

Because of the divergent terms in (\ref{eq:coulomb_expand}) the diagonalization
in the limit $\mathbf{k\rightarrow0}$ is not trivial. The first eigenvector
$\mathbf{E}_{1}^{\mathbf{k}}$ corresponding to the divergent eigenvalue
$v_{1}(\mathbf{k})=4\pi/k^{2}$ is, however, easy to obtain from the
analytic projection of $e^{i\mathbf{k}\cdot\mathbf{r}}/\sqrt{V}$
on the biorthogonal mixed-product-basis functions \begin{equation}
E_{1I}^{\mathbf{k}}=\frac{1}{\sqrt{V}}\int_{V}\tilde{M}_{I}^{\mathbf{k}^{{\scriptstyle *}}}(\mathbf{r})\, e^{i\mathbf{k\cdot r}}\, d^{3}r=\left\{ \begin{array}{ll}
c_{I\mathbf{0}}^{*}(\mathbf{k})=\frac{4\pi i^{L}}{\sqrt{\Omega}}Y_{LM}^{*}(\mathbf{e_{k}})\int_{0}^{s_{a}}r^{2}j_{L}(kr)M_{aLP}(r)\, dr & \textrm{ for }I=(aLMP)\,,\\
\delta_{\mathbf{G0}} & \textrm{ for }I=\mathbf{G}\,,\end{array}\right.\label{eq:constant_eigenvector}\end{equation}
which in the limit $\mathbf{k}\rightarrow\mathbf{0}$ becomes \begin{equation}
E_{1I}^{\mathbf{0}}=\left\{ \begin{array}{ll}
\sqrt{4\pi s_{a}^{3}/(3\Omega)} & \textrm{ for }I=(a001)\,,\\
\delta_{\mathbf{G0}} & \textrm{ for }I=\mathbf{G}\,,\\
0 & \textrm{ otherwise }.\end{array}\right.\end{equation}
Here we have assumed that the constant MT function of atom $a$ is
normalized and identified with the index $(a001)$. The other eigenvectors
$\mathbf{E}_{\mu}^{\mathbf{0}}$ and eigenvalues $v_{\mu}(\mathbf{0})$
for $\mu>1$ are obtained by diagonalizing the last term of the formal
expansion (\ref{eq:coulomb_expand_PW})\begin{equation}
\overline{v}_{IJ}=\sum_{\mathbf{G}\ne\mathbf{0}}c_{I\mathbf{G}}^{*}c_{J\mathbf{G}}\frac{4\pi}{G^{2}}\,,\end{equation}
which is unknown so far. The matrix $v_{IJ}^{(0)}$, calculated in
the previous section, contains $\overline{v}_{IJ}$ but also the spherical
average of the second square bracket in (\ref{eq:coulomb_expand_PW}).
If we denote this spherical average by $w_{IJ}$, then we can write\begin{equation}
\overline{v}_{IJ}=v_{IJ}^{(0)}-w_{IJ}\,.\end{equation}
 In order to evaluate $w_{IJ}$ we introduce the natural basis\begin{subequations}\label{eqall:NaturalBasis}\begin{eqnarray}
k_{-1} & = & \frac{1}{\sqrt{2}}(k_{x}-ik_{y}),\quad\quad\,\, k_{1}=\frac{1}{\sqrt{2}}(-k_{x}-ik_{y}),\quad\quad\,\, k_{0}=k_{z}\,,\\
\partial_{-1} & = & \frac{1}{\sqrt{2}}\left(\partial_{x}+i\partial_{y}\right),\quad\partial_{1}=\frac{1}{\sqrt{2}}\left(-\partial_{x}+i\partial_{y}\right),\quad\partial_{0}=\partial_{z}\,,\end{eqnarray}
\end{subequations}which allows us to write the \textbf{k}-dependent
terms in the second bracket of (\ref{eq:coulomb_expand_PW}) in terms
of spherical harmonics according to\begin{eqnarray}
\mathbf{e}_{\mathbf{k}}^{\mathrm{T}}\Delta c_{J\mathbf{0}}\mathbf{e}_{\mathbf{k}} & = & \frac{4\pi}{3}\sum_{m=-1}^{1}\sum_{m'=-1}^{1}Y_{1m}^{*}(\mathbf{e_{k}})Y_{1m'}(\mathbf{e_{k}})\partial_{m}^{*}\partial_{m'}c_{J\mathbf{0}}\\
 & = & \frac{1}{3}\sum_{m=-1}^{1}\partial_{m}^{*}\partial_{m}c_{J\mathbf{0}}+\frac{4\pi}{3}\sum_{m=-1}^{1}\sum_{m'=-1}^{1}C_{1m'1m2(m-m')}Y_{2(m-m')}^{*}(\mathbf{e_{k}})\partial_{m}^{*}\partial_{m'}c_{J\mathbf{0}}\,,\nonumber \\
\left(\mathbf{e}_{\mathbf{k}}\cdot\nabla c_{I\mathbf{0}}^{*}\right)\left(\mathbf{e}_{\mathbf{k}}\cdot\nabla c_{J\mathbf{0}}\right) & = & \frac{4\pi}{3}\sum_{m=-1}^{1}\sum_{m'=-1}^{1}Y_{1m}^{*}(\mathbf{e_{k}})Y_{1m'}(\mathbf{e_{k}})\partial_{m}^{*}c_{I\mathbf{0}}^{*}\partial_{m'}c_{J\mathbf{0}}\\
 & = & \frac{1}{3}\sum_{m=-1}^{1}\partial_{m}^{*}c_{I\mathbf{0}}^{*}\partial_{m}c_{J\mathbf{0}}+\frac{4\pi}{3}\sum_{m=-1}^{1}\sum_{m'=-1}^{1}C_{1m'1m2(m-m')}Y_{2(m-m')}^{*}(\mathbf{e_{k}})\partial_{m}^{*}c_{I\mathbf{0}}^{*}\partial_{m'}c_{J\mathbf{0}}\,,\nonumber \end{eqnarray}
where we define $\partial_{m}c_{I\mathbf{0}}=\left.\partial_{m}c_{I\mathbf{0}}(\mathbf{k})\right|_{\mathbf{k=0}}$
and similar abbreviations. The last equation follows from the identity
\begin{equation}
\mathbf{e}_{\mathbf{k}}\cdot\nabla c_{I\mathbf{0}}=\sqrt{\frac{4\pi}{3}}\sum_{m=-1}^{1}Y_{1m}(\mathbf{e_{k}})\partial_{m}c_{I\mathbf{0}}\,.\end{equation}
When we take the spherical average, the harmonics with $l>0$ vanish,
and we obtain\begin{equation}
w_{IJ}=\frac{4\pi}{3}\sum_{m=-1}^{1}\left[\left(\partial_{m}^{*}c_{I\mathbf{0}}^{*}\right)\left(\partial_{m}c_{J\mathbf{0}}\right)+\frac{1}{2}c_{I\mathbf{0}}^{*}\partial_{m}^{*}\partial_{m}c_{J\mathbf{0}}+\frac{1}{2}c_{J\mathbf{0}}\partial_{m}\partial_{m}^{*}c_{I\mathbf{0}}^{*}\right]\end{equation}
with\begin{eqnarray}
c_{I\mathbf{0}} & = & \left\{ \begin{array}{ll}
\sqrt{4\pi s_{a}^{3}/(3\Omega)} & \textrm{ for }I=(a001)\,,\\
\Theta_{\mathbf{-G}} & \textrm{ for }I=\mathbf{G}\,,\\
0 & \textrm{ otherwise}\,,\end{array}\right.\\
\partial_{m}c_{I\mathbf{0}} & = & \left\{ \begin{array}{ll}
-\delta_{L1}\delta_{Mm}\sqrt{\frac{4\pi}{3\Omega}}i\int_{0}^{s_{a}}r^{3}M_{a1P}(r)\, dr & \textrm{ for }I=(a1MP)\,,\\
0 & \textrm{ otherwise}\,,\end{array}\right.\\
\sum_{m=-1}^{1}\partial_{m}\partial_{m}^{*}c_{I\mathbf{0}} & = & \left\{ \begin{array}{ll}
-\delta_{L0}\sqrt{\frac{4\pi}{\Omega}}\int_{0}^{s_{a}}r^{4}M_{a0P}(r)\, dr & \textrm{ for }I=(a0MP)\,,\\
0 & \textrm{ otherwise}\,.\end{array}\right.\end{eqnarray}

\section{Test calculations\label{sec:test}}

\begin{figure}[b]
\includegraphics[%
  width=0.35\textwidth,
  angle=-90]{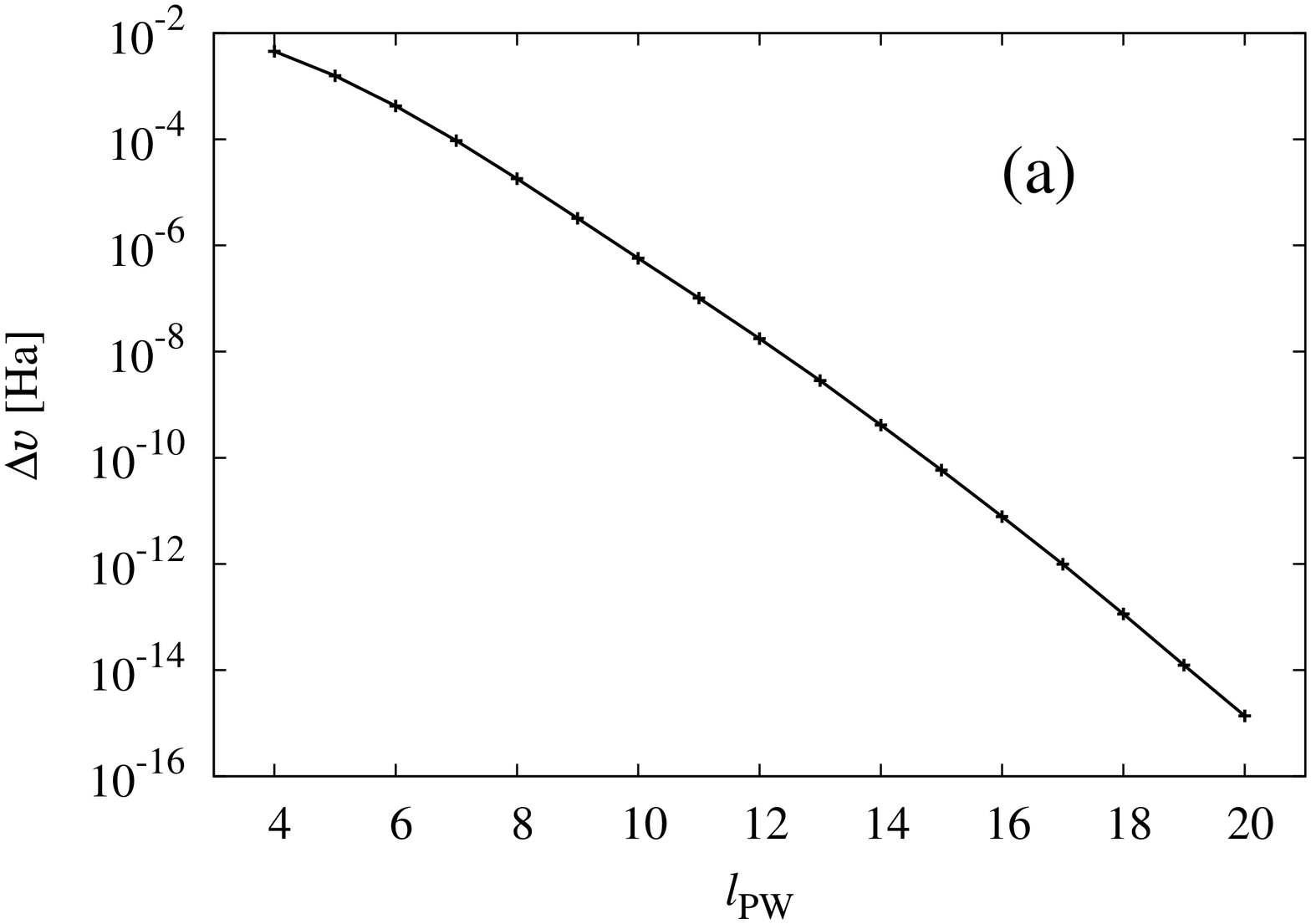}\includegraphics[%
  width=0.35\textwidth,
  angle=-90]{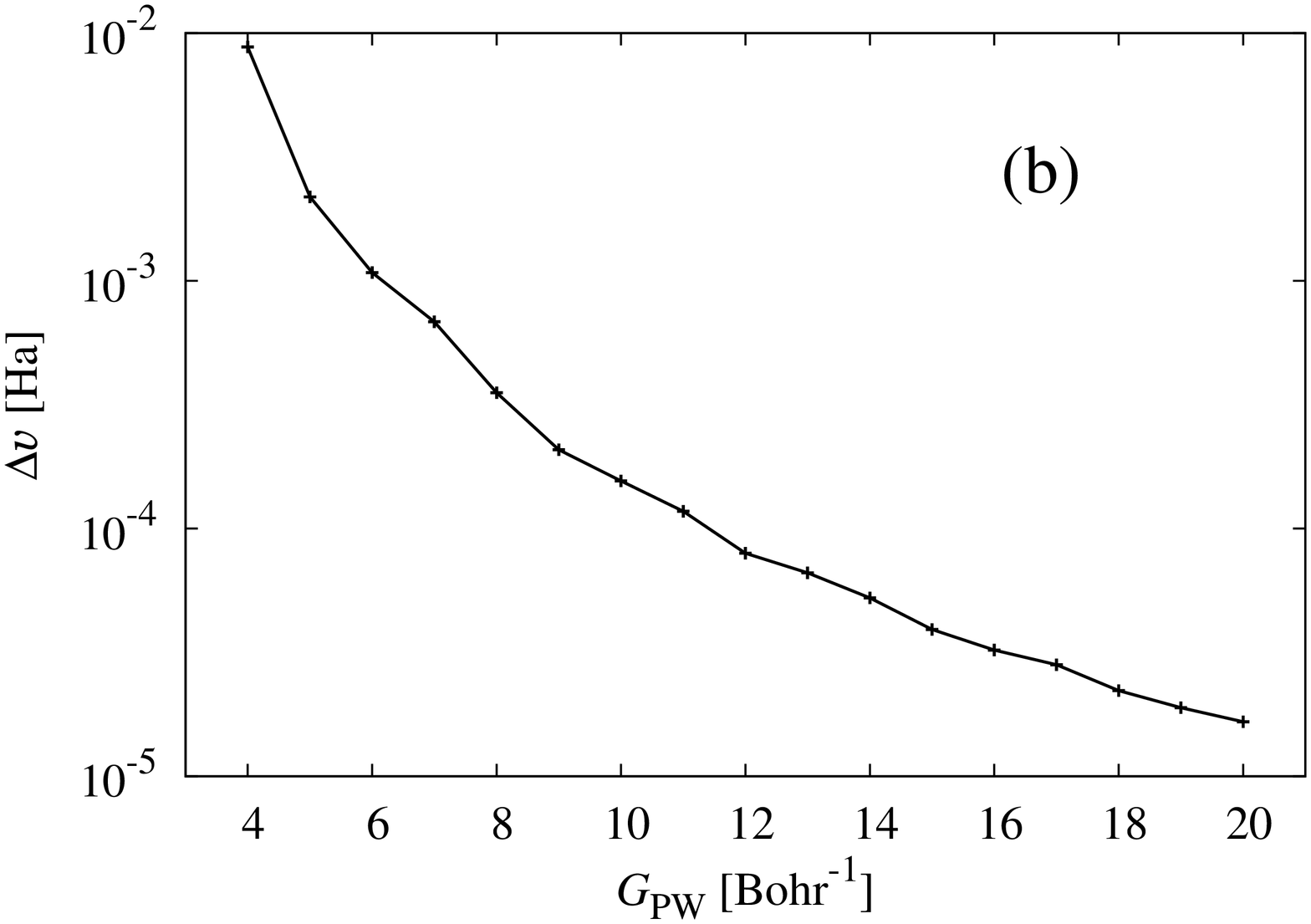}

\caption{\label{cap:convergence}Average root mean square deviation $\Delta v$
from the converged matrix elements (MT-IPW and IPW-IPW) as functions
of (a) the convergence parameters $l_{\mathrm{PW}}$ and (b) the reciprocal
cutoff radius $G_{\mathrm{PW}}$ for the Fourier transform of the
step function in (\ref{eq:IPW_Gexpand}). The mixed product basis
was optimized for Si bulk.}
\end{figure}

Apart from the evaluation of (\ref{eq:struct}), which is easily converged
to high precision by means of the Ewald summation technique, and the
radial meshes for numerical integration inside the MT spheres, the
cutoff value $l_{\mathrm{PW}}$ is the only convergence parameter
in the construction of the Coulomb matrix elements presented above.
On the other hand, in an alternative implementation that uses the
representation (\ref{eq:IPW_Gexpand}) rather than the Rayleigh expansion
for the IPWs the matrix elements must be converged with respect to
the reciprocal cutoff radius $G_{\mathrm{PW}}$. Figure~\ref{cap:convergence}
compares the convergence behavior of these two approaches. The curves
indicate the root mean square deviation of the Coulomb matrix for
bulk silicon, averaged over all matrix elements MT-IPW and IPW-IPW
and over 64 $\mathbf{k}$ points, from the fully converged results
calculated with $l_{\mathrm{PW}}=26$. In both cases we employ the
same cutoff parameters $G_{\mathrm{max}}^{\prime}=3.6\,\mathrm{Bohr}^{-1}$
and $L_{\mathrm{max}}=4$ for the mixed product basis, the MT functions
are constructed from products $u_{al0}^{\sigma}(r)u_{al'0}^{\sigma}(r)$
with $l\le2$ and $l'\le3$. On average this yields 411 basis functions
per $\mathbf{k}$ point. It is evident that the results obtained with
the present method converge much faster than those obtained with the
Fourier transform of the step function. Furthermore, at the same level
of accuracy the present method is typically by a factor of 10--100
faster.%

\begin{figure}[t]
\includegraphics[%
  scale=0.65,
  angle=-90]{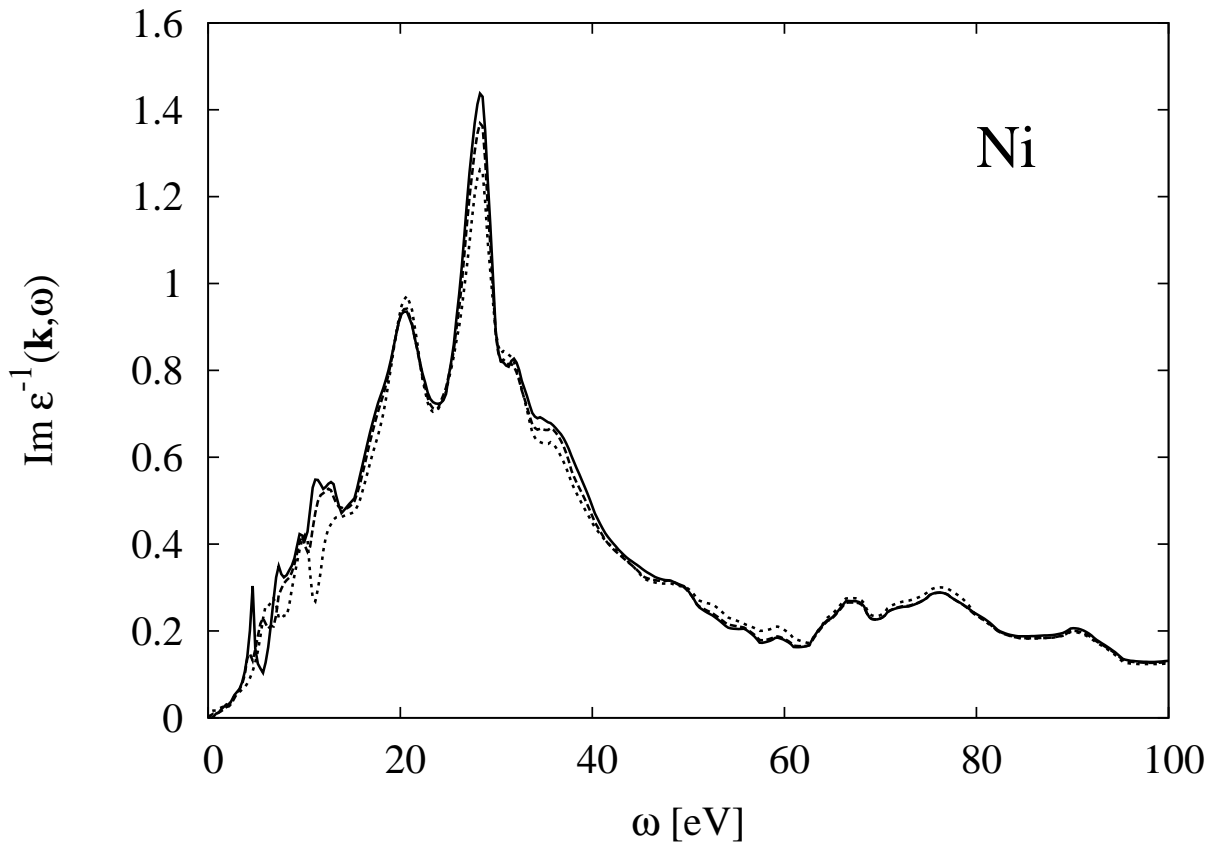}

\caption{\label{cap:Ni_EELS}EELS spectra of spin-polarized Ni for $\mathbf{k}=2\pi/a_{\mathrm{Ni}}(\xi,\xi,\xi)$
with $\xi=0$ (solid line), $\xi=0.1$ (dashed line), and $\xi=0.2$
(dotted line).}
\end{figure}
As an application we now consider the simulation of experimental spectroscopies
related to the complex dielectric function, which describes many-body
screening effects in a correlated electron system. In electron-energy-loss
spectroscopy (EELS), for example, the measured differential scattering
cross section is directly proportional to the imaginary part of a
diagonal element of the inverse dielectric function \cite{Onida2002}\begin{equation}
\varepsilon^{-1}(\mathbf{k},\omega)=\frac{1}{V}\iint\varepsilon^{-1}(\mathbf{r},\mathbf{r}';\omega)e^{i\mathbf{k}\cdot(\mathbf{r}'-\mathbf{r})}d^{3}r\, d^{3}r'\,,\label{eq:epsilon_EELS}\end{equation}
whereas in optical absorption one measures the imaginary part of \cite{Adler1962}\begin{equation}
\varepsilon_{\mathrm{M}}(\omega)=\lim_{\mathbf{k}\rightarrow\mathbf{0}}1/\varepsilon_{\mathrm{M}}^{-1}(\mathbf{k},\omega)\,.\label{eq:epsilon_optic}\end{equation}
In the framework of many-body perturbation theory the dielectric function
is written as\begin{equation}
\varepsilon(\mathbf{r},\mathbf{r}';\omega)=\delta(\mathbf{r}-\mathbf{r}')-\int v(\mathbf{r},\mathbf{r}'')P(\mathbf{r}'',\mathbf{r}';\omega)\, d^{3}r''\label{eq:epsilon_MBPT}\end{equation}
with the polarization function $P(\mathbf{r},\mathbf{r}';\omega)$
and the Coulomb interaction $v(\mathbf{r},\mathbf{r}')=1/|\mathbf{r}-\mathbf{r}'|$.
We use the random-phase approximation 

\begin{eqnarray}
P(\mathbf{r},\mathbf{r}';\omega) & = & \sum_{\sigma}\sum_{n,\mathbf{q}}^{\mathrm{occ}}\sum_{n',\mathbf{k}}^{\mathrm{unocc}}\varphi_{n\mathbf{k}}^{\sigma^{{\scriptstyle *}}}(\mathbf{r})\varphi_{n'\mathbf{q+k}}^{\sigma}(\mathbf{r})\varphi_{n\mathbf{k}}^{\sigma}(\mathbf{r}')\varphi_{n'\mathbf{q+k}}^{\sigma^{{\scriptstyle *}}}(\mathbf{r}')\label{eq:RPA}\\
 &  & \times\left(\frac{1}{\omega+\epsilon_{n\mathbf{k}}^{\sigma}-\epsilon_{n'\mathbf{q+k}}^{\sigma}+i\eta}-\frac{1}{\omega-\epsilon_{n\mathbf{k}}^{\sigma}+\epsilon_{n'\mathbf{q+k}}^{\sigma}-i\eta}\right)\,,\nonumber \end{eqnarray}
where $\eta$ is a positive infinitesimal. As $P(\mathbf{r},\mathbf{r}';\omega)$
contains products of wave functions evaluated at $\mathbf{r}$ and
$\mathbf{r}'$, it can be represented in the mixed product basis as\begin{equation}
P(\mathbf{r},\mathbf{r}';\omega)=\sum_{I,J}\int_{\mathrm{BZ}}P_{IJ}(\mathbf{k},\omega)M_{I}^{\mathbf{k}}(\mathbf{r})M_{J}^{\mathbf{k}^{{\scriptstyle *}}}(\mathbf{r}')\, d^{3}k\end{equation}
with complex coefficients\begin{equation}
P_{IJ}(\mathbf{k},\omega)=\iint P(\mathbf{r},\mathbf{r}';\omega)\tilde{M}_{I}^{\mathbf{k}^{{\scriptstyle *}}}(\mathbf{r})\tilde{M}_{J}^{\mathbf{k}}(\mathbf{r}')\, d^{3}r\, d^{3}r'\,.\end{equation}
Next we transform this matrix to the basis given by (\ref{eq:basis_eigenvec}).
This yields $P_{\mu\nu}(\mathbf{k},\omega)$, which in the limit $\mathbf{k}\rightarrow\mathbf{0}$
decomposes into head, wing, and body elements as discussed in Section~\ref{sub:Diagonalization}.
We use the tetrahedron method for integrations over the BZ. 

In the long-wave-length limit $\mathbf{k}\rightarrow\mathbf{0}$ we
must carefully expand the polarization function around $\mathbf{k}=\mathbf{0}$,
since it is multiplied with $v(\mathbf{r},\mathbf{r}')$ in (\ref{eq:epsilon_MBPT}),
which diverges in this limit. Because of the orthogonality of the
wave functions the projection $\langle E_{1}^{\mathbf{k}}\varphi_{n\mathbf{q}}^{\sigma}|\varphi_{n'\mathbf{q+k}}^{\sigma}\rangle=\langle e^{i\mathbf{k\cdot r}}\varphi_{n\mathbf{q}}^{\sigma}|\varphi_{n'\mathbf{q+k}}^{\sigma}\rangle/\sqrt{V}$
is linear in lowest order in $\mathbf{k}$ for interband transitions
with $n\neq n'$. We calculate this leading term with $\mathbf{k}\cdot\mathbf{p}$
perturbation theory \cite{Baroni1986}. For a metallic system the
sum in (\ref{eq:RPA}) also contains contributions from intraband
transitions with $n=n'$ at $\mathbf{k}=\mathbf{0}$. It can be shown
that these are nonzero only for the head element and given analytically
by the Drude formula \cite{Ziesche}, which is quadratic in $\mathbf{k}$.
The latter depends on the plasma frequency, which we obtain by an
integration over the Fermi surface. In conclusion, the head and wing
elements of $P_{\mu\nu}(\mathbf{k},\omega)$ are quadratic and linear
in $\mathbf{k}$, respectively. If we use the symmetrized dielectric
matrix\begin{equation}
\tilde{\varepsilon}_{\mu\nu}(\mathbf{k},\omega)=\delta_{\mu\nu}-v_{\mu}^{1/2}(\mathbf{k})P_{\mu\nu}(\mathbf{k},\omega)v_{\nu}^{1/2}(\mathbf{k})\,,\end{equation}
where the $v_{\mu}(\mathbf{k})$ are the eigenvalues of $v_{IJ}(\mathbf{k})$,
then all elements of $\tilde{\varepsilon}_{\mu\nu}(\mathbf{k},\omega)$
are finite because $v_{1}^{1/2}(\mathbf{k})=\sqrt{4\pi}/k$. We note
that the diagonal quantities considered above remain unchanged with
this symmetrized definition. As the first eigenvector of $v_{IJ}(\mathbf{k})$
corresponds to the projection of $e^{i\mathbf{k\cdot r}}/\sqrt{V}$
onto the biorthogonal mixed product basis, the head element $\tilde{\varepsilon}_{11}^{-1}(\mathbf{k},\omega)$
directly equals the spectroscopic function (\ref{eq:epsilon_EELS}). 

In figure \ref{cap:Ni_EELS} we show EELS spectra $\textrm{Im}\,\varepsilon^{-1}(\mathbf{k},\omega)$
for spin-polarized Ni at three $\mathbf{k}$ vectors $2\pi/a_{\mathrm{Ni}}(\xi,\xi,\xi)$
with $\xi=0.0,\,0.1,\,0.2$ and the lattice constant $a_{\mathrm{Ni}}=6.66\,\mathrm{Bohr}$
We use the parameters $l_{\mathrm{max}}=8$, $G_{\mathrm{max}}=3.57\,\mathrm{Bohr}^{-1}$
for the FLAPW and $L_{\mathrm{max}}=4$, $G_{\mathrm{max}}=5.00\,\mathrm{\mathrm{Bohr}^{-1}}$
for the mixed product basis. The BZ is sampled by 1661 points in its
irreducible wedge, corresponding to a 40$\times$40$\times$40 $\mathbf{k}$-point
mesh in the full zone. As the spectrum extends over a wide energy
range up to 100~eV, we augment the FLAPW basis by second and third
energy derivatives as local orbitals to guarantee an accurate description
of high-lying conduction states \cite{Friedrich2006}. For the construction
of the MT functions $M_{aLP}(r)$ we employ products $u_{alp}^{\sigma}(r)u_{al'p}^{\sigma}(r)$
with $l\le2$, $l'\le3$, and $p=0$, i.e., energy derivatives ($p\ge1$)
are neglected. In the calculation of (\ref{eq:RPA}) we take 118 conduction
and the 10 valence states as well as the eight $3s$ and $3p$ core
states into account. As we invert the dielectric function, local-field
effects are fully included. As seen from the figure, the spectra are
very similar for the three $\mathbf{k}$ vectors. When compared with
the curves calculated at finite $\mathbf{k}$ points, the spectrum
for $\mathbf{k}=\mathbf{0}$ clearly constitutes the limit $\mathbf{k}\rightarrow\mathbf{0}$. 

\begin{figure}[t]
\includegraphics[%
  scale=0.65,
  angle=-90]{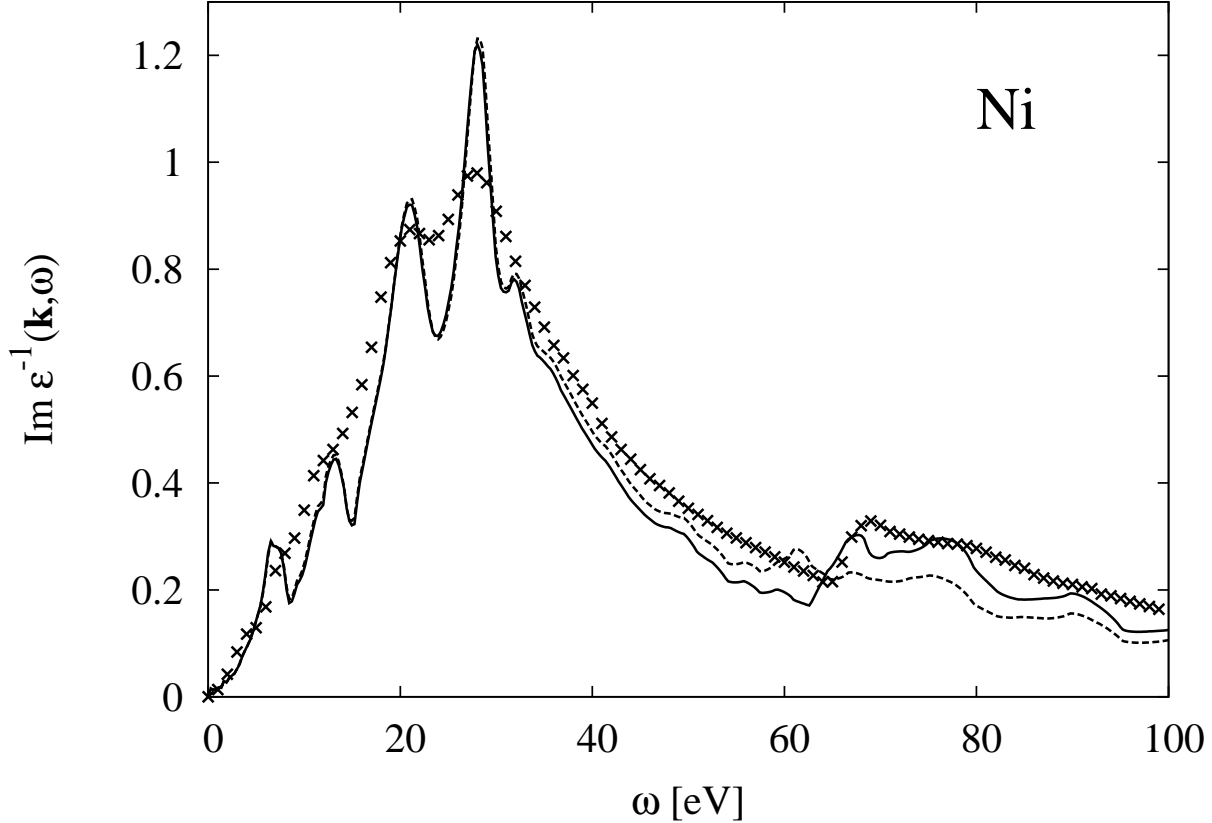}

\caption{\label{cap:Ni_EELS_2}EELS spectra of spin-polarized Ni for $\mathbf{k}=2\pi/a_{\mathrm{Ni}}(0.25,0,0)$
with (solid line) and without (dashed line) core-state contributions.
The inclusion of transitions from core into conduction states gives
rise to a shallow peak starting at 62~eV, which is also seen in experiment
(symbols) \cite{Feldkamp1979}.}
\end{figure}
As already pointed out, the spectra in figure~\ref{cap:Ni_EELS}
already include transitions from the $3s$ and $3p$ core states into
conduction states. Figure \ref{cap:Ni_EELS_2} shows a comparison
of spectra calculated with (solid line) and without (dashed line)
these core-state contributions at $\mathbf{k}=2\pi/a_{\mathrm{Ni}}(0.25,0,0)$.
The largest difference between the two curves is seen around 62~eV,
which corresponds roughly to the threshold energy required to excite
a $3p$ electron above the Fermi level. These additional transitions
give rise to a shallow peak, which is also observed in experiments
with an onset at the same energy. The inclusion of transitions from
$3s$ and $3p$ core states into conduction states within our all-electron
method thus brings the calculated spectrum very close to experiment
(symbols) \cite{Feldkamp1979}.

\section{Summary\label{sec:Summary} }

In this work we have derived formulas for the Coulomb matrix elements
within the all-electron FLAPW method. As the Coulomb interaction couples
two incoming and two outgoing states, a suitable basis set must be
capable of accurately represent wave-function products. Such a set
is given by the mixed product basis, which contains MT functions as
well as interstitial plane waves. We use the Rayleigh expansion for
the latter, because it makes a very efficient numerical implementation
possible. Furthermore, we have derived an exact expansion of the Coulomb
matrix around $\mathbf{k}=\mathbf{0}$ that isolates all divergent
terms $\sim k^{-2}$ and $\sim k^{-1}$. Most of these vanish if we
then make a basis transformation to the eigenvectors of the Coulomb
matrix. The properties of this new basis set are formally similar
to those of a plane-wave basis. In particular, response functions
decompose into head, wing, and body elements with the same characteristic
dependence on $\mathbf{k}$. However, the basis construction of this
involves no approximation, and the accuracy of the FLAPW basis set
is completely preserved.

As an illustration we have shown EELS spectra for ferromagnetic Ni
at different $\mathbf{k}$ vectors including $\mathbf{k}=\mathbf{0}$.
Very good agreement with experiment was achieved over a large energy
window by taking core-electron contributions into account.

\begin{ack}
We gratefully acknowledge financial support from the Deutsche Forschungsgemeinschaft
through the Priority Program 1145.
\end{ack}
\appendix

\section{Mathematical relations\label{sec:Used-relations}}

In the derivations of Sections \ref{sec:Coulomb} and \ref{sec:Expansion}
we have used the following relations:\begin{eqnarray}
j_{l-1}(x)+j_{l+1}(x) & = & (2l+1)j_{l}(x)/x\,,\label{eq:recursion_bessel}\\
\frac{d}{dx}j_{l}(x) & = & \frac{l}{x}j_{l}(x)-j_{l+1}(x)\,,\label{eq:sphes_deriv}\\
\frac{d}{dx}j_{l}(x) & = & j_{l-1}(x)-\frac{l+1}{x}j_{l}(x)\,,\label{eq:sphes_deriv2}\\
j_{l}(x) & = & \frac{x^{l}}{(2l+1)!!}\left(1-\frac{x^{2}}{4l+6}+\mathrm{O}(x^{4})\right)\,,\label{eq:sphes_expand}\\
\mathbf{e_{k}\cdot e_{G}} & = & \frac{4\pi}{3}\sum_{m=-1}^{1}Y_{1m}(\mathbf{e_{k}})Y_{1m}^{*}(\mathbf{e_{G}}),\label{eq:kG_Y1Y1}\\
Y_{1m}(\mathbf{e_{a}})Y_{1m'}^{*}(\mathbf{e_{a}}) & = & \frac{1}{4\pi}\delta_{mm'}+C_{1m1m'2(m'-m)}Y_{2,m'-m}^{*}(\mathbf{e_{a}})\,,\label{eq:Y1+Y1}\\
Y_{1m}^{*}(\mathbf{e_{k+G}}) & = & Y_{1m}^{*}(\mathbf{e_{G}})+\frac{2k}{3G}\left[Y_{1m}^{*}(\mathbf{e_{k}})-2\pi\sum_{m'=-1}^{1}C_{m'm}Y_{2,m-m'}^{*}(\mathbf{e_{G}})Y_{1m'}^{*}(\mathbf{e_{k}})\right]+\mathrm{O}(k^{2})\label{eq:Y1_expand}\\
(\mathbf{e_{k}\cdot e_{G}})^{2} & = & \frac{1}{3}+\frac{8\pi}{15}\sum_{m'=-2}^{2}Y_{2m'}^{*}(\mathbf{e_{k}})Y_{2m'}(\mathbf{e_{G}})\,,\label{eq:kG2_}\\
Y_{1m}(\mathbf{e_{k}}) & = & \sqrt{\frac{3}{4\pi}}\frac{k_{m}}{k}\,,\label{eq:Y1_}\end{eqnarray}
\begin{equation}
\sum_{m=-1}^{1}\partial_{m}\partial_{m}^{*}f(k)Y_{lm}(\mathbf{e}_{\mathbf{k}})=\frac{1}{k^{2}}Y_{lm}(\mathbf{e}_{\mathbf{k}})\left[\partial_{k}\left(k^{2}\partial_{k}\right)-l(l+1)\right]f(k)\,.\label{eq:SUMDmDm}\end{equation}

\section{Integrals over spherical Bessel functions\label{sec:Integrals-sphbes}}

The derivations in Section \ref{sec:Coulomb} give rise to a number
of integrals over spherical Bessel functions that can be evaluated
analytically. Explicit formulas for (\ref{eq:integral_sphbes_1})
follow from the recursion relations (\ref{eq:recursion_bessel})-(\ref{eq:sphes_deriv2})
\begin{equation}
{\cal I}_{l}(q,r)=\left\{ \begin{array}{ll}
\frac{r^{l+2}}{q}j_{l+1}(qr) & \textrm{ if }q\neq0\,,\\
\delta_{l0}\frac{1}{3}r^{3} & \textrm{ if }q=0\,,\end{array}\right.\label{eq:sphbes_prim1}\end{equation}
\begin{equation}
{\cal J}_{al}(q,r)=\left\{ \begin{array}{ll}
\frac{1}{q}\left[\frac{1}{r^{l-1}}j_{l-1}(qr)-\frac{1}{s_{a}^{l-1}}j_{l-1}(qs_{a})\right] & \textrm{ if }q\neq0\,,\\
\delta_{l0}\frac{1}{2}\left(s_{a}^{2}-r^{2}\right) & \textrm{ if }q=0\,.\end{array}\right.\label{eq:sphbes_prim2}\end{equation}
We can also find an analytic expression for the double integral (\ref{eq:integral_sphbes_2}),
because the above integration formulas and the recursion relation
(\ref{eq:recursion_bessel}) lead to the solution\begin{subequations}\begin{eqnarray}
{\cal K}_{al}(q,q') & = & \frac{2l+1}{q'^{2}}\int_{0}^{s_{a}}r^{2}j_{l}(qr)j_{l}(q'r)\, dr-\frac{s_{a}^{3}}{qq'}j_{l+1}(qs_{a})j_{l-1}(q's_{a})\nonumber \\
 & = & \frac{s_{a}^{3}}{q^{2}-q'^{2}}\left[\frac{q'}{q}j_{l+1}(qs_{a})j_{l-1}(q's_{a})-\frac{q}{q'}j_{l-1}(qs_{a})j_{l+1}(q's_{a})\right]\label{eq:sphbessel_integral1}\\
 & = & s_{a}^{3}\left[\frac{j_{l+1}(qs_{a})j_{l+1}(q's_{a})}{qq'}+\frac{2l+1}{2l+3}\frac{j_{l+2}(qs_{a})j_{l}(q's_{a})-j_{l}(qs_{a})j_{l+2}(q's_{a})}{q^{2}-q'^{2}}\right]\,,\label{eq:sphbessel_integral2}\end{eqnarray}
\end{subequations}where we used the symmetry of ${\cal K}_{al}(q,q')$
with respect to $q$ and $q'$ to eliminate $\int_{0}^{s_{a}}r^{2}j_{l}(qr)j_{l}(q'r)dr$.
The expressions (\ref{eq:sphbessel_integral1}) and (\ref{eq:sphbessel_integral2})
are stable for large and small $q$, $q'$, respectively. The limiting
cases are\begin{subequations}\begin{eqnarray}
\lim_{q'\rightarrow0}{\cal K}_{al}(q,q') & = & \delta_{l0}\frac{s_{a}^{3}}{3q^{2}}\left[qs_{a}j_{1}(qs_{a})+j_{2}(qs_{a})\right]\quad\textrm{for }q\ne0\quad,\\
\lim_{q'\rightarrow q}{\cal K}_{al}(q,q') & = & \frac{s_{a}^{3}}{2q^{2}}\left[(2l+3)j_{l+1}^{2}(qs_{a})-(2l+1)j_{l}(qs_{a})j_{l+2}(qs_{a})\right]\quad\textrm{for }q\neq0\quad,\\
\lim_{q,q'\rightarrow0}{\cal K}_{al}(q,q') & = & \delta_{l0}\frac{2}{15}s_{a}^{5}\quad.\end{eqnarray}
\end{subequations}

\end{document}